\documentclass[10pt,a4paper, twoside]{article}
\usepackage[dvips]{graphicx}
\usepackage{amssymb,amsmath}
\usepackage{natbib}
\usepackage{multirow}
\usepackage{subfigure}
\usepackage{rotating}
\usepackage{setspace}
\usepackage[usenames]{color}
\usepackage[left=2cm,top=2cm,bottom=2cm]{geometry}
\begin{document}

\noindent This manuscript version is distributed under the CC-BY-NC-ND (Creative Commons) license.\\

\noindent It has appeared in final form as:\\
Van Albada SJ, Gray RT, Drysdale PM, Robinson PA. Mean-field modeling of the basal
ganglia-thalamocortical system. II. Dynamics of parkinsonian oscillations (2009) J
Theor Biol 257:664--688, DOI: 10.1016/j.jtbi.2008.12.013\\

\newpage

\title{Mean-field modeling of the basal ganglia-thalamocortical system. II. Dynamics of parkinsonian oscillations} 
\author{S. J. van Albada$^{1,2}$ \and R. T. Gray $^{1,2,3}$ \and P. M. Drysdale $^{1,2}$ \and P. A. Robinson$^{1,2,4}$}
\maketitle

\begin{center}
$^1$School of Physics, The University of Sydney\\
New South Wales 2006, Australia\\
$^2$The Brain Dynamics Centre, Westmead Millennium Institute\\
Westmead Hospital and Western Clinical School of the University of Sydney\\
Westmead, New South Wales 2145, Australia\\
$^3$National Centre in HIV Epidemiology and Clinical Research\\
Faculty of Medicine, The University of New South Wales\\
Sydney, New South Wales 2010, Australia\\
$^4$ Faculty of Medicine, The University of Sydney\\
New South Wales 2006, Australia
\end{center}


\begin{abstract}
Neuronal correlates of Parkinson's disease (PD) include a shift to lower frequencies in the electroencephalogram (EEG) and enhanced synchronized oscillations at 3--7 and 7--30 Hz in the basal ganglia, thalamus, and cortex. This study describes the dynamics of a recent physiologically-based mean-field model of the basal ganglia-thalamocortical system, and shows how it accounts for many key electrophysiological correlates of PD. Its detailed functional connectivity comprises partially segregated direct and indirect pathways through two populations of striatal neurons, a hyperdirect pathway involving a corticosubthalamic projection, thalamostriatal feedback, and local inhibition in striatum and external pallidum (GPe). In a companion paper, realistic steady-state firing rates were obtained for the healthy state, and after dopamine loss modeled by weaker direct and stronger indirect pathways, reduced intrapallidal inhibition, lower firing thresholds of the GPe and subthalamic nucleus (STN), a stronger projection from striatum to GPe, and weaker cortical interactions. Here it is shown that oscillations around 5 and 20 Hz can arise with a strong indirect pathway, which also causes increased synchronization throughout the basal ganglia. Furthermore, increased theta power with progressive nigrostriatal degeneration is correlated with reduced alpha power and peak frequency, in agreement with empirical results. Unlike the hyperdirect pathway, the indirect pathway sustains oscillations with phase relationships that coincide with those found experimentally. Alterations in the responses of basal ganglia to transient stimuli accord with experimental observations. Reduced cortical gains due to both nigrostriatal and mesocortical dopamine loss lead to slower changes in cortical activity and may be related to bradykinesia. Finally, increased EEG power found in some studies may be partly explained by a lower effective GPe firing threshold, reduced GPe-GPe inhibition, and/or weaker intracortical connections in parkinsonian patients. Strict separation of the direct and indirect pathways is not necessary to obtain these results.
\end{abstract}

\section{Introduction}

\citet{Parkinson1817} described a syndrome with symptoms including a stooped posture, shuffling gait (festination), sleep disturbances, and rest tremor. This disorder, which also leads to slowness of movement (bradykinesia), difficulty initiating movements (akinesia), and rigidity, was subsequently called Parkinson's disease (PD). It is one of the most common movement disorders, affecting 0.5--3\% of those over 65 \citep{Tanner1996}. The pathological hallmark of PD is the progressive degeneration of dopaminergic neurons in the substantia nigra pars compacta (SNc) and to a lesser extent the ventral tegmental area (VTA) \citep{Bernheimer1973, Ehringer1960, Hirsch1988, Uhl1985}. These nuclei provide dopaminergic input to the basal ganglia, a group of gray matter structures close to the thalamus concerned with reinforcement learning and the facilitation and modulation of movement \citep{Graybiel1990, Mink1996}. The main structures comprising the basal ganglia are the striatum, the substantia nigra, the globus pallidus internal (GPi) and external (GPe) segments, and the subthalamic nucleus (STN). Alterations in the associative and limbic functions of the basal ganglia are responsible for some of the cognitive symptoms and mood disturbances seen in PD.

In a companion paper [\citet{vanAlbada2008}; henceforth referred to as Paper I] we introduced a physiologically realistic mean-field model of the basal ganglia-thalamocortical system (BGTCS), and assessed changes in average firing rates caused by loss of nigrostriatal dopamine. We found that an increase in the strength of cortical transmission to striatal cells expressing the D2 class of dopamine receptor, with or without a concurrent decrease in the connection strength to D1-expressing cells, could account for the majority of changes in firing rates observed in parkinsonism. Reduced lateral inhibition in the GPe, a lower GPe firing threshold, and reduced intracortical inhibition helped account for the reported lack of change in cortical rate \citep{Goldberg2002}. Lower STN and GPe thresholds combined with decreased intrapallidal inhibition explained the relatively large increase in STN rate and relatively small decrease in GPe rate observed experimentally \citep{Bergman1994,Boraud1998,Filion1991,Goldberg2002,Heimer2002,Hutchison1994,Kreiss1997,Pan1988,Walters2007}. Increases in GPe rate due to changes in GPe and STN firing thresholds and intrapallidal inhibition were limited by stronger striatal inhibition of the GPe, which is expected with dopamine loss \citep{Floran1997, Querejeta2001}. 

The purpose of this paper is to analyze the dynamics of the model presented in Paper I in the healthy and parkinsonian states, and to compare modeling results with experimental findings on electrophysiological changes caused by dopaminergic denervation. Dynamical changes with dopamine loss include altered responses to transient stimuli, a lower frequency of the alpha peak and increased relative low-frequency power in the electroencephalogram (EEG), and synchronized oscillations around 5 and 20 Hz throughout the BGTCS. The present work is devoted to modeling these electrophysiological changes, which are described in detail in Sec. \ref{sec:experimental_overview}. Using parameters that accord well with known physiology, we obtain not only realistic firing rates (see Paper I), but also realistic EEG spectra and responses to transient stimuli, oscillations in the theta and beta ranges, and enhanced synchronization in the basal ganglia. In Sec. \ref{sec:neuronal_substrates} we review the relevant functional anatomy and electrophysiological changes found experimentally, and place our model in context by describing possible origins of parkinsonian oscillations. Section \ref{sec:dynamics} details the model equations and parameter changes in PD, followed by an analysis of changes in neuronal responsiveness in Sec. \ref{sec:responses}. Oscillations and frequency spectra predicted by the model are discussed in Sec. \ref{sec:oscillations}. 

\section{Physiological background}\label{sec:neuronal_substrates}

In Sec. \ref{sec:connectivity} we summarize the functional connections of the BGTCS included in the model, which were described more fully in Paper I. Section \ref{sec:experimental_overview} provides an overview of the electrophysiological changes caused by nigrostriatal degeneration. Possible neuronal substrates of parkinsonian oscillations are discussed in Sec. \ref{sec:origins}.

\subsection{Functional connectivity}\label{sec:connectivity}

The main functional connections of the BGTCS are depicted in Fig. \ref{fig:hypotheses}. Excitatory input from the cortex reaches the basal ganglia mainly at the striatum, of which 90--95\% of cells are medium spiny projection neurons \citep{Kemp1971b}. Medium spiny neurons are classified according to their predominant type of dopamine receptor (D1 or D2). Although a proportion of striatal neurons has both D1 and D2 receptors, a partial segregation appears to exist between these populations \citep{Hersch1995, Inase1997, LeMoine1995, Lester1993, Meador-Woodruff1991}. A number of studies have suggested that dopaminergic input from the SNc affects D1 and D2-expressing cells differently, primarily increasing the effect of cortical input on D1 cells, and primarily decreasing the sensitivity of D2 cells to cortical input \citep{Gerfen1990, Mallet2006}. Projections from the SNc thus effectively modulate the synaptic strengths of corticothalamic inputs, but are not considered as a dynamic part of the model.

Both D1 and D2 striatal neurons exert an inhibitory influence on their projection sites: D1 neurons on the GPi and the substantia nigra pars reticulata (SNr), and D2 neurons on the GPe. SNr and GPi are the main output nuclei of the basal ganglia, sending inhibitory projections primarily to the ventral anterior (VA) and ventrolateral (VL) thalamic nuclei \citep{Parent1990, Parent1995}, but also to the centromedian-parafascicular complex (CM-Pf) \citep{Kim1976, Parent2001}. The pathway cortex-D1-GPi/SNr-thalamus, where the loop is completed via excitatory (glutamatergic) projections to the cortex, is termed the \emph{direct pathway} \citep{Albin1989,Alexander1990}. Since D1 inhibits the output nuclei, which in turn inhibit the thalamus, the direct pathway as a whole is excitatory.

The GPe sends an important inhibitory projection to the STN, which in turn excites the output nuclei. Therefore, the pathway cortex-D2-GPe-STN-GPi/SNr-thalamus as a whole is inhibitory, so that cortical activation results in inactivation of the thalamus. This pathway is referred to as the \emph{indirect pathway} \citep{Albin1989,Alexander1990}. Projections from GPe to STN are reciprocated by excitatory STN-GPe projections. There also exists an important \emph{hyperdirect pathway} from the cerebral cortex to the STN, particularly from the frontal lobe \citep{Afsharpour1985, Hartmann1978,Nambu2000, Parent1995b}. The direct, indirect, and hyperdirect pathways are illustrated in Fig. \ref{fig:hypotheses}(a).

Significant thalamostriatal projections arise from the relay nuclei CM-Pf, VA, and VL \citep{Carpenter1981, Gonzalo2002, Parent1990, Sadikot1990, Sadikot1992}. Empirical results indicate that the associated neurotransmitter is glutamate \citep{Haber2004, Sadikot1992}. The GPe contains a dense network of local axon collaterals \citep{Kita1994,Nambu1997,Ogura2000}, and sends a projection to the GPi \citep{Hazrati1990,Sato2000b,Shink1995,Smith1994}. GABAergic striatal interneurons receiving input from the cortex powerfully inhibit medium spiny neurons, which also provide local axon collaterals \citep{Bolam2000,Koos1999,Somogyi1981,Wilson2007}. Although striatal axon collaterals appear to exert excitatory effects at hyperpolarized membrane potentials, their influence becomes inhibitory near spike threshold \citep{Plenz2003,Taverna2004}. Since the strength of local interactions increases with the firing rate, and in view of the strong inhibitory action of striatal interneurons, we model intrastriatal connections as inhibitory.

Reciprocal connections exist between the relay nuclei and the thalamic reticular nucleus (TRN), both of which receive excitatory (glutamatergic) input from the cerebral cortex. The relay nuclei project back to the cortex, and the cortex contains populations of short-range inhibitory interneurons and long-range excitatory (pyramidal) cells. Sensory input reaches the thalamic relay nuclei from the brainstem via glutamatergic and cholinergic afferents. In contrast, the basal ganglia do not receive significant projections from ascending sensory pathways \citep{Elble2002}.

Anatomical and physiological studies have shown that connections in the BGTCS form three mostly separate circuits (sensorimotor, association, and limbic), which are further organized into somatotopic regions \citep{Alexander1986, Alexander1990}. Since the sensorimotor circuit is most relevant to parkinsonian motor symptoms, in the present work we ignore projections to and from the amygdala, dorsal raphe nucleus, hippocampus, and pedunculopontine nucleus (PPN) of the brainstem, which mostly affect limbic territories. We do not treat the remaining territories separately for two reasons. First, since many empirical studies do not distinguish between sensorimotor, association, and limbic pathways, or between pathways corresponding to different somatotopic regions, sufficient physiological data are not available for each circuit separately. Second, the similarity of the connectivity patterns of all circuits suggests that oscillations may be generated by a common mechanism. 

\subsection{Overview of experimental findings}\label{sec:experimental_overview}

The wide-ranging effects of nigrostriatal dopamine depletion on the dynamics of the BGTCS include alterations of responses to transient stimuli or during volitional tasks. Regional cerebral bloodflow appears to be diminished in the supplementary motor area (SMA) and the dorsolateral prefronal cortex, whereas other cortical areas display increased activation during motor tasks \citep{Jenkins1992, Playford1992,Sabatini2000}. On the other hand, cortical activation has been shown to be suppressed specifically during tasks with significant involvement of the caudate nucleus, but increased otherwise \citep{Monchi2007}. The increased activation of cortical regions in the absence of striatal involvement may reflect compensation for deficits caused by nigrostriatal dopamine loss \citep{Samuel1997}, or degeneration of direct mesocortical dopaminergic afferents \citep{Mattay2002}. The responsiveness of striatal neurons is expected to change depending on the type of dopamine receptor they primarily express (D1 or D2). Dopamine potentiates the activation of D1-expressing neurons by glutamate unless they are in a hyperpolarized state \citep{Cepeda1998,Hernandez-Lopez1997,Kiyatkin1996,Nicola2000}. In contrast, the responsiveness to glutamate is suppressed via D2 receptor activation \citep{Hsu1995, Levine1996, Toan1985, Umemiya1997}. GPi neurons display more widespread and vigorous responses to passive limb movements in parkinsonian monkeys \citep{Bergman1994,Filion1988}, especially to extension torque \citep{Wichmann1994}. \citet{Zold2007} reported an increase in the number of excitatory responses in the globus pallidus (GP; the rodent homolog of GPe) of rats with moderate nigrostriatal damage, but inhibitory responses increased after more extensive damage. A reduction in pallidal activity upon cortical stimulation was also observed in a rat model of PD \citep{Magill2001}. The STN responds more vigorously to cortical stimulation in parkinsonian rats \citep{Magill2001}. Furthermore, the duration and magnitude of both excitatory and inhibitory responses in the STN of African green monkeys increased after SNc lesion with the neurotoxin 1-methyl-4-phenyl-1,2,3,6-tetrahydropyridine (MPTP), most cells increasing their firing rate \citep{Bergman1994}. 

EEG and magnetoencephalographic (MEG) spectra show that relative power in the delta and theta bands (i.e., the proportions of the total power in these frequency bands) is increased in PD patients compared to age-matched controls \citep{Bosboom2006, Neufeld1994, Stoffers2007}. In non-demented patients, increased delta and theta power is associated with more severe motor and cognitive symptoms \citep{Bosboom2006, Neufeld1994}, and may be specific to patients with REM sleep behavior disorder \citep{Gagnon2004}. PD is also accompanied by lower alpha peak frequencies, especially in patients with dementia \citep{Sinanovic2005,Soikkeli1991}; e.g., \citet{Soikkeli1991} reported mean alpha peak frequencies of 9.6 Hz in controls, 8.3 Hz in non-demented PD patients, and 6.8 Hz in demented patients. Some studies also reported that absolute EEG power is increased across the spectrum in PD patients by an amount independent of medication state \citep{Moazami-Goudarzi2008, Tanaka2000}, although \citet{Yaar1983} found that levodopa increased spectral power in the left occipital lobe over all frequency bands. Increased corticothalamic coherence at $\sim$7 Hz \citep{Sarnthein2007} is in line with greater EEG power and a lower frequency of the alpha peak. Overall, EEG power appears to decline with advancing dementia, but according to \citet{Tanaka2000} demented patients still had higher delta (1.5--6 Hz), theta (6.5--8 Hz), and beta (13.5--30 Hz) power than controls. On the other hand, \citet{Soikkeli1991} reported higher delta and theta power, but lower beta power in demented patients than in control subjects. EEG frequency reduction in PD may be partly caused by changes in non-dopaminergic neurotransmitter systems \citep{Stoffers2007}. Especially the cholinergic and noradrenergic systems are implicated, since demented PD patients have reduced cortical cholinergic activity due to degeneration of the nucleus basalis of Meynert \citep{Candy1983,Dubois1983}, as well as more significant noradrenergic deficits than non-demented patients \citep{Cash1987}. Moreover, loss of cholinergic afferentation from the basal forebrain has been related to EEG frequency reduction with age \citep{Buzsaki1988, Longo1966, Metherate1995}. We propose a mechanism by which nigrostriatal denervation itself also contributes to EEG frequency reduction in PD.

PD is also associated with enhanced oscillations at 3--7 Hz and 7--30 Hz in the BGTCS, which have been observed in humans and in animal models of PD in the GPi and GPe \citep{Filion1991, Levy2002, Nini1995, Raz2000}, STN \citep{Bergman1994, Levy2000, Wang2005}, SNr \citep{Wichmann1999}, pallidal and cerebellar-receiving areas of the thalamus \citep{Lamarre1979, Lenz1994, Magnin2000,Ohye1974,Ohye1989}, striatal medium spiny neurons \citep{Dejean2008} and tonically active interneurons \citep{Raz2001}, and sensory and motor cortices \citep{Alberts1969, Cordeau1960, Lamarre1979, Timmermann2003, Volkmann1996}. In the GPi of PD patients, $\sim$5 Hz oscillations were found to be more common than 15--20 Hz oscillations \citep{Levy2001}. A percentage of GPe and GPi neurons in monkeys treated with MPTP displayed 5--8 Hz oscillations for short periods during tremor, and approximately 15 Hz oscillations in both pallidal segments in the absence of tremor, which sometimes persisted over longer intervals in the GPi \citep{Filion1991}. Microelectrode recordings revealed stable 11--30 Hz synchronization and more transient $\sim$5 Hz coherence between pairs of neurons in the STN of PD patients, particularly those with tremor \citep{Levy2000}. The large power of beta oscillations in the STN \citep{Kuhn2005}, and the differential modulation of low- and high-frequency activity by dopaminergic medication and during periods of intermittent tremor \citep{Silberstein2003, Wang2005}, suggest that $\sim$20 Hz oscillations are not generated as a harmonic of the $\sim$5 Hz rhythm.

\subsection{Possible origins of oscillations}\label{sec:origins}

Despite the similar frequencies of slow oscillations in the BGTCS and resting tremor, it is unclear whether a direct causal relationship exists between the two, since rhythmically discharging cells have been found in the basal ganglia even without obvious tremor \citep{Bergman1994, Levy2001, Ohye1974, Wichmann1994b, Williams2002}, as has a lack of coherence even during tremor \citep{Bergman1994}. Furthermore, synchronized beta oscillations are also more prominent in patients with intermittent tremor than in non-tremulous patients, indicating an association between higher-frequency rhythms and tremor \citep{Levy2000} (although we will see that relatively strong beta resonances can occur in a system that is likely to support theta oscillations, which may well be more directly related to tremor). The fact that even random electrical \citep{Alberts1972} or 10--30 Hz magnetic stimulation \citep{Topka1999} of the motor cortex can induce 4--7 Hz tremor also shows that basal ganglia oscillations need not directly determine tremor frequency. Finally, the basal ganglia do not appear to initiate movements under normal conditions \citep{Horak1984,Mink1996,Rivlin-Etzion2006}.

Involvement of the cerebellum in the generation of parkinsonian tremor has been suggested since tremor-like oscillations in the cerebellar-receiving ventralis intermedius (Vim) nucleus of the thalamus are stronger and show greater coherence with the EMG than oscillations in basal ganglia targets \citep{Lenz1988}, and Vim has long been a preferred surgical target for PD \citep{Okun2004}. However, all forms of tremor are accompanied by abnormal cerebellar activation, likely due to proprioceptive feedback from the limbs \citep{Deuschl2001}. More telling is the study by \citet{Deuschl1999} on a patient who developed PD 17 years after a stable lesion of the right cerebellar hemisphere. This patient exhibited a bilateral tremor with a frequency of 3.1 Hz on the side of the cerebellar lesion, and 4.3 Hz on the intact side, demonstrating that an intact olivocerebellar circuit is not necessary to produce tremor. The tremor on the lesioned side was a combination of resting, postural, and intention tremors, and was interpreted as rubral or Holmes tremor. The fact that levodopa is effective against Holmes tremor indicates that it is due to a combination of cerebellar and nigrostriatal damage, which is confirmed by PET imaging \citep{Remy1995,Velez2002}. The cerebellum may prevent nigrostriatal degeneration from causing tremor during voluntary movement, and may modulate the frequency of resting tremor \citep{Deuschl1999}. The small degree of entrainment and phase resetting of parkinsonian tremor that can be achieved with imposed periodic movements \citep{Rack1986} and mechanical stimulation \citep{Lee1981}, respectively, also indicates the limited influence of peripheral factors. Even if the frequency or presence of tremor is not directly determined by oscillations in the BGTCS, an important role for the basal ganglia in the production of tremor is therefore evident. Furthermore, a basal ganglia contribution to akinesia, bradykinesia, and rigidity is implied by the fact that pallidal or STN lesions or high-frequency stimulation can ameliorate these symptoms \citep{Gross1997, Iacono1995, Limousin1998, Meissner2005}.

\begin{figure}[htp]
\centering
  \includegraphics[width=420pt]{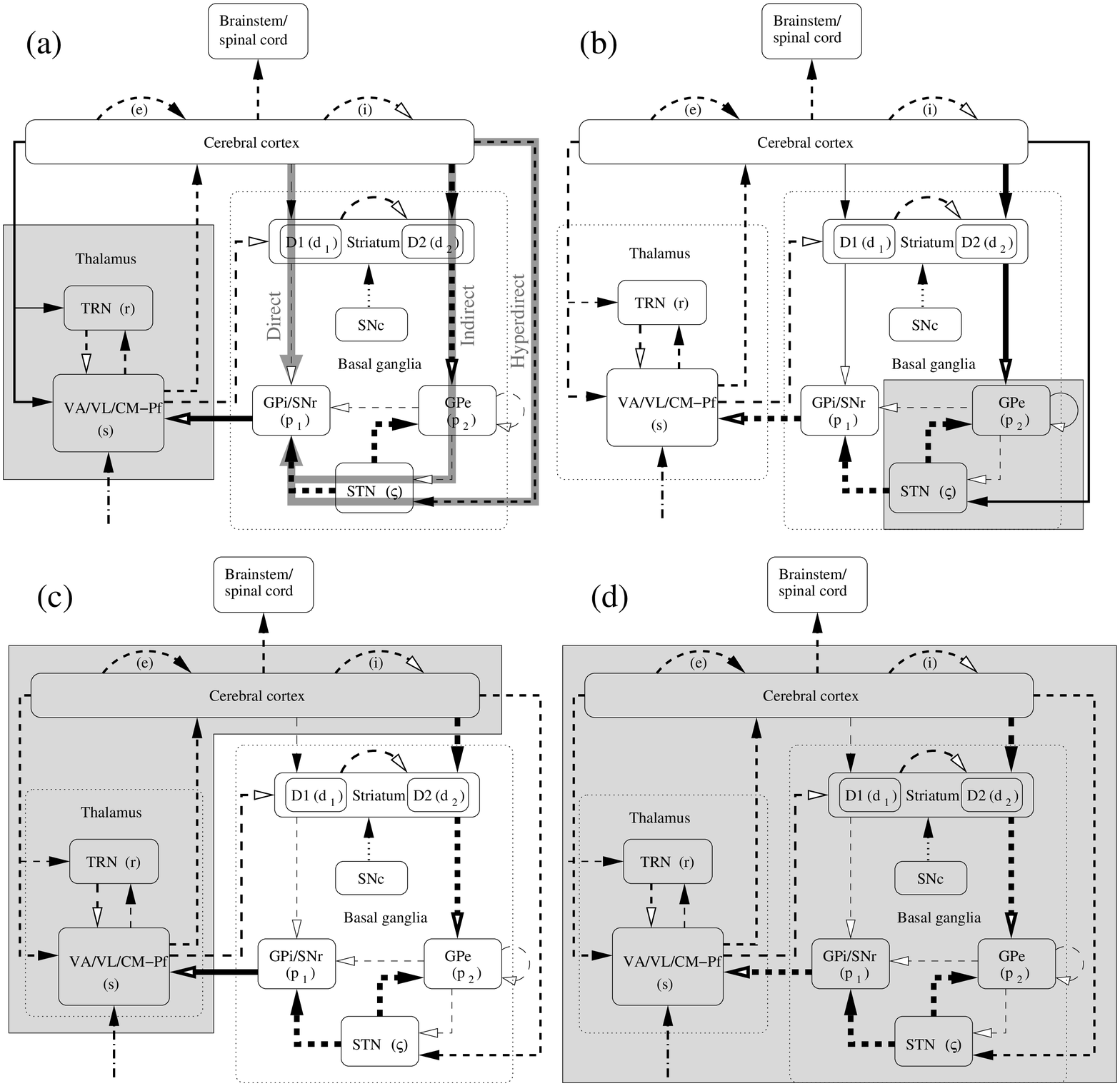}
  \caption{Four possible network origins of parkinsonian tremor, indicated by the shaded areas. Signals are ultimately transmitted to the muscles via the corticospinal tract. Filled arrowheads indicate excitatory projections, and open arrowheads inhibitory ones, with the thickness of the lines representing the strength of connections with respect to the normal state. Dashed lines indicate the transfer of tremor-related activity. Although some tremor activity may be transferred via the remaining projections, arrows are drawn solid to emphasize that the receiving structures can generate oscillations without rhythmic input. Subscripts used for the various populations are given in parentheses. (a) Thalamic relay nuclei are hyperpolarized by the GPi, causing low-threshold calcium spike bursts. These bursts become synchronized through intrathalamic interactions. Gray arrows indicate the direct, indirect, and hyperdirect pathways. (b) Increased striatal input to GPe and decreased intrapallidal inhibition cause rhythmic oscillations in the GPe-STN system. Rhythmic bursts are seen in the thalamus due to periodic input from the GPi. (c) Resonances arise in the corticothalamic loop or intracortically after hypoactivation of the thalamic relay nuclei. (d) Resonances arise in the cortico-basal ganglia-thalamic circuit as a whole. Combinations of mechanisms are possible; for instance, rhythmic activity generated in the thalamus could be enhanced within the STN-GPe network, or both cortico-basal ganglia-thalamic and STN-GPe circuits could act as sources of rhythmic oscillations. } 
\label{fig:hypotheses}
\end{figure}

Due to the widespread nature of the oscillatory activity in PD, various parts of the BGTCS are implicated as possible sources \citep{Deuschl2000,Elble1996}. Several hypotheses have been put forward concerning the origin of parkinsonian oscillations, which can be grouped into the following four---not necessarily mutually exclusive---categories for our purposes.\\

\noindent \emph{Hypothesis of thalamic origin}

The first scenario, illustrated in Fig. \ref{fig:hypotheses}(a), places the origin in the thalamus \citep{Hurtado1999, Llinas1984, Pare1990, Sarnthein2007}. The thalamus is implicated since striatal dopamine depletion leads to disinhibition of GPi and SNr, increasing inhibition of the VA and VL thalamic nuclei \citep{Hutchison1994, Magnin2000, Pan1988}. Hyperpolarization of thalamic relay neurons has been shown to cause low-threshold calcium spike bursts \citep{Deschenes1982, Jeanmonod1996, Llinas1982, Llinas2006}, some of which occur rhythmically at an interburst frequency of $\sim$4 Hz in patients with symptoms related to thalamic hypoactivation, including parkinsonian tremor \citep{Jeanmonod1996,Ohye1974}. Such rebound bursts in response to tonic overinhibition by the GPi and SNr may be synchronized by lateral inhibition within the TRN, and their rhythmicity may be partly determined by network interactions between the TRN and the relay nuclei \citep{Buzsaki1991}. The thalamic oscillations would mediate their effect on cortical and muscular activity through projections to the prefrontal and premotor cortices \citep{Sarnthein2007}. Rhythmic oscillations may be relayed to the basal ganglia either via the striatum or via the corticosubthalamic projection, and enhanced within the loop formed by STN and GPe \citep{Baufreton2005}, which appear to become more sensitive to rhythmic inputs following loss of striatal and extrastriatal dopamine \citep{Bevan2006}. Finally, GPe and STN could output rhythmic oscillations via projections to GPi and SNr.

Until the recent finding of 5--13 Hz oscillations in medium spiny neurons of unanesthetized rats with nigrostriatal lesions \citep{Dejean2008}, evidence for the involvement of the corticostriatal and striatopallidal projections was mostly indirect. The delay in obtaining this experimental evidence is explained by the low firing rate of medium spiny neurons, which prevents autocorrelation functions from showing prominent oscillations. \citet{Dejean2008} overcame this problem by using peri-event histograms, and focused on high-voltage spindles to track rhythmic activity in the unanesthetized condition. An earlier argument put forward for the involvement of striatal projections \citep{Murer2002} is the relatively long phase delay (20--30 ms) between rhythmic cortical activity at and activity in the STN and GPi of PD patients off medication \citep{Marsden2001, Williams2002}, which is longer than the corticosubthalamic delay of $<$10 ms estimated in a few studies \citep{Ashby2001,Nambu2000}. Further indirect evidence for striatal transmission of tremor-like rhythms is that the membrane potentials of striatal medium spiny neurons reliably follow $\sim$5 Hz cortical rhythms induced by light anesthesia in animals \citep{Mahon2001, Wilson1993}. In addition, \citet{Tseng2001} demonstrated that slow ($\sim$1 Hz) rhythmic cortical inputs induce oscillations at the same frequency in striatal neurons of anesthetized rats with nigrostriatal lesions. Oscillatory activity at $\sim$15 Hz is also observed in the tonically active cholinergic interneurons of the rat striatum \citep{Raz1996, Raz2001}, although the relationship between these oscillations and the activity of medium spiny neurons is unclear. The cholinergic interneurons may oscillate in response to nigrostriatal afferents, input from the GPe \citep{Sato2000b}, or input from CM-Pf \citep{Lapper1992}. 

On the other hand, involvement of the corticosubthalamic pathway is implied by the fact that the discharge rate and pattern of GP neurons is not significantly affected by cortical ablation in anesthetized rats, whereas in the STN, this abolishes slow oscillations coherent with the EEG \citep{Magill2001}. The regularization of GP \citep{Ni2000}, SNr \citep{Burbaud1995, Tseng2000} and GPi \citep{Wichmann1994} activity upon STN lesion has also been put forward as evidence for the contribution of the corticosubthalamic projection, but these findings do not exclude the possibilities that oscillatory activity arises in the STN-GPe network or in cortico-basal ganglia-thalamic loops involving striatopallidal projections. 

A possible objection to the hypothesis of a thalamic pacemaker for parkinsonian tremor is the fact that different body parts often tremble at slightly different frequencies, suggesting the absence of the overall synchronizing influence of the TRN \citep{Pare1990}. However, we have seen above that cortical and peripheral factors appear to play a role in determining tremor frequency. \citet{Lenz1993} and \citet{Zirh1998} noted a difference between the pattern of interspike intervals in thalamic bursts in parkinsonism and the gradual lengthening of interspike intervals characteristic of calcium spike-associated bursts. This may indicate that rhythmic GPi inputs to slightly depolarized thalamic neurons, rather than increased tonic GPi activity and consequent low-threshold calcium spike bursts, are responsible for the bursting patterns observed in the thalamus.\\

\noindent \emph{Hypothesis of origin in STN-GPe loop}

A second hypothesis, illustrated in Fig. \ref{fig:hypotheses}(b), states that changes in tonic input to the STN-GPe system cause these nuclei to produce oscillatory activity, which spreads to the basal ganglia output nuclei, thalamus, and cortex \citep{Deuschl2000,Terman2002}. It has been shown that STN neurons can switch from a single-spike mode to a burst-firing mode upon hyperpolarization by the GPe \citep{Beurrier1999, Bevan2000}. Although the GPe generally becomes less active in parkinsonism, excitation by the STN may induce periods of enhanced GPe firing, periodically hyperpolarizing STN neurons and causing rebound bursts.
This mechanism was suggested by \citet{Plenz1999}, who showed that the rat STN-GP network undergoes spontaneous 0.4--1.8 Hz oscillations  in the absence of dopamine. 
\citet{Terman2002} modeled networks of STN and GPe neurons and showed that increased synaptic input from the striatum to the GPe and/or weakened intrapallidal inhibition can lead to slow ($< 1$ Hz) synchronized oscillations in these nuclei or clustered rhythms at 4--6 Hz, depending on the network architecture and the STN-GPe connection strength. The reduction in lateral inhibition within the GPe would be caused by the enhanced release of enkephalin \citep{Stanford1999, Steiner1998}, preventing desynchronization of GPe neurons by intranuclear interactions. Oscillations in the STN-GP circuit have also been replicated in computational models by \citet{Gillies2002,Humphries2001,Humphries2006}.

A requirement of this hypothesis is that the STN-GPe network be able to generate rhythmic activity in the absence of rhythmically modulated input. Both an experimental study \citep{Magill2001}, and a modeling study \citep{Humphries2006} of rat basal ganglia have shown that small subpopulations of GP and STN neurons can sustain rhythmic oscillations around 1 Hz. However, \citet{Magill2001} reported that most correlated bursting in GPe and SNr was abolished by cortical desynchronization or ablation. A few other studies have also suggested that the isolated STN-GPe network does not generate significant rhythmic bursting activity.
First, a study in rats showed that GABA antagonists, mimicking reduced pallidal input to STN, can make bursting patterns more marked, but cannot change an irregular spiking mode into a bursting mode \citep{Urbain2002}. Furthermore, \citet{Wilson2006} showed that rhythmic oscillations were absent in a slice preparation of the isolated pallidosubthalamic network from dopamine-depleted mice. \citet{Wilson2006} pointed out that inputs from striatum and/or cortex may be needed to induce synchronized oscillations in these nuclei in the slice preparation, although these inputs were not necessary in the rat cell culture \citep{Plenz1999}. However, the slice preparation of \citet{Wilson2006} may not have left sufficient GP projections intact to significantly inhibit the STN and evoke rebound firing in that nucleus. \\

\noindent \emph{Hypothesis of origin in corticothalamic loops}

A third possibility, depicted in Fig. \ref{fig:hypotheses}(c), is that rhythmic activity arises from reverberations within the corticothalamic network. If we assume that the frequency of resting tremor is directly determined by central oscillations, the lack of coherence between GPi cells firing near tremor frequencies and the tremor EMG \citep{Lemstra1999} appears to be evidence for a thalamic or corticothalamic source of tremor oscillations. \citet{Magnin2000} reported a larger proportion of tremor-locked cells in the GPi than in the thalamus, which provides evidence both for and against this hypothesis. On the one hand, it suggests that oscillations arise upstream from the thalamus; on the other hand, rhythmic oscillations may not be relayed directly from GPi to thalamus. Although \citet{Magnin2000} found no synchronization between individual pallidal-receiving thalamic cells and tremor, it is possible that the average activity of a larger subpopulation was modulated rhythmically in a coherent manner, since even weak correlations between individual cells can lead to strong coherence at the population level \citep{Schneidman2006}.\\

\noindent \emph{Hypothesis of origin in basal ganglia-thalamocortical loops}

Finally, resonances may originate in the BGTCS as a whole, rather than being confined to any particular region \citep{Deuschl2000, Wichmann2003} [Fig. \ref{fig:hypotheses}(d)]. This hypothesis implies that the rhythmicity would be determined by the delay for signals to complete a full loop from cortex through the basal ganglia, thalamus, and back to the cortex. Such a mechanism was proposed by \citet{Leblois2006}, who described a neuronal network model of interacting direct and hyperdirect pathways, of which the latter sustained oscillations at $\sim$11 Hz when regulation by the direct pathway was suppressed. The objection that GPi and thalamic activity appear to be non-coherent during tremor \citep{Lemstra1999, Magnin2000} can be raised against this hypothesis as well as against the hypothesis that oscillations arise in the STN-GPe network. However, as discussed above, lack of coherence of individual cells with the EMG or with other rhythmic cells does not appear to rule out the involvement of these cells in the generation of parkinsonian oscillations. 

Our paper investigates which of the above hypotheses is supported by the known functional anatomy of the BGTCS if the dynamics is governed by mean-field activity.

\section{Dynamical equations}\label{sec:dynamics}

The form of the model developed in Paper I and used here is based on \citet{Rennie1999} and \citet{Robinson1997, Robinson1998b, Robinson1998, Robinson2002, Robinson2005}, work that partly built on models developed by \citet{Freeman1975}, \citet{Jirsa1996}, \citet{Nunez1974, Nunez1995}, \citet{Wright1996}, and others. The model incorporates synaptic and dendritic integration effects, nonlinear response functions, axonal conduction delays, and spreading of wave-like activity along the cortical surface. The basic model equations were largely given in Paper I, but are summarized in Sec. \ref{sec:basic} for convenience. Section \ref{sec:linear} contains the equations that govern small perturbations about fixed point. In Sec. \ref{sec:changes} we describe the parameter changes used to model parkinsonism.

\subsection{Basic equations}\label{sec:basic}

The neuronal populations of the model will be indicated by the following subscripts: $e$, cortical excitatory; $i$, cortical inhibitory; $d_1$, striatal D1 cells; $d_2$, striatal D2 cells; $p_1$, GPi/SNr; $p_2$, GPe; $\varsigma$, STN; $s$, thalamic relay nuclei; and $r$, TRN. We refer to brainstem input using a subscript $n$. The subscript $s$ indicates both specific relay nuclei and the diffusely projecting CM-Pf complex. 
The dependence of the mean firing rate $Q_a(\mathbf{r},t)$ of each population of neurons $a$ on the cell-body potential $V_a(\mathrm{r}, t)$ relative to resting is modeled by a sigmoidal function taking the form \citep{Robinson2002}
\begin{equation}\label{eq:1}
Q_a(\mathbf{r},t) \equiv S_a[V_a(\mathbf{r},t)]= \frac{Q_a^{\mathrm{max}}}{1+\mathrm{exp}[-(V_a(\mathbf{r},t) - \theta_a)/\sigma']}. 
\end{equation}
In previous works the cortex was modeled as two-dimensional since it has a large surface area but is comparatively thin, whereas other components were treated as uniform. In the present work we indicate the spatial coordinate $\mathbf{r}$ for completeness, although we will only consider spatially uniform solutions. The functional dependence (\ref{eq:1}) results from averaging the response functions of neurons with slightly different firing thresholds. The quantity $Q_a^{\mathrm{max}}$ is the maximum firing rate, $\theta_a$ is the mean threshold potential, and $\sigma'\pi/\sqrt{3}$ is the standard deviation of firing thresholds. The latter is taken to be equal for all components, since we lack precise knowledge of the ranges of firing thresholds in different populations. 

Changes in the cell-body potential of type $a$ neurons are triggered by pulses that arrive from type $b$ neurons after an axonal delay $\tau_{ab}$, are filtered by the dendritic tree, and summed at the cell body. The effect of an incoming pulse rate  $\phi_b(\mathbf{r},t-\tau_{ab})$ on the cell-body potential depends on the connection strengths between the neural populations. These are made up of the mean number of synapses, $N_{ab}$, and the typical time-integrated change in cell-body potential per incoming pulse, $s_{ab}$. Connection strengths are thus given by the products $\nu_{ab} = N_{ab}s_{ab}$. The change in the average cell-body potential of type $a$ neurons due to the summation and temporal integration of incoming signals becomes \citep{Robinson2004}
\begin{align}
D_{\alpha\beta}(t)V_a(\mathbf{r},t) = \sum_b \nu_{ab}\phi_b(\mathbf{r},t-\tau_{ab}),\label{eq:dend1}\\
D_{\alpha\beta}(t) = \frac{1}{\alpha\beta}\frac{d^2}{dt^2}+\left(\frac{1}{\alpha} + \frac{1}{\beta}\right)\frac{d}{dt} + 1.\label{eq:dend2}
\end{align}
The differential operator $D_{\alpha\beta}(t)$ approximates filtering of signals by the synapses and the dendritic tree \citep{Rennie2000, Robinson1997}. We assume $\alpha < \beta$ without loss of generality, which in practice means that $\alpha$ is the decay rate and $\beta$ the rise rate of the cell-body potential. The synapses and dendrites reduce the power of signals for angular frequency $\omega \gtrsim \alpha$ and especially for $\omega \gtrsim \beta$. In more complicated models, $\alpha$ and $\beta$ may be taken to depend on both the sending and receiving populations, and the relevant neurotransmitter \citep{Rennie2000}. 

The cortical signal depends not only on resonances with underlying structures, but also on corticocortical interactions. Many experiments have revealed propagating waves of neuronal activity upon local cortical stimulation \citep{Chervin1988,Nunez1974,Schiff2007,Xu2007}, a feature previously included in a number of modeling studies \citep{Bressloff2001,Bressloff2003,Jirsa1996,Jirsa1997,Nunez1995}. Although \citet{Nunez1995} focused on standing waves obtained when cortical waves are weakly damped, physiological and modeling evidence indicates that considerable damping occurs, and should be taken into account \citep{Robinson2001b,Robinson2004,Wright1995}. We thus model propagation effects along the cortical surface due to long-range excitatory connections via a damped-wave equation with source term $Q_e(\mathbf{r}, t)$ \citep{Robinson1997, Robinson2001b}, 
\begin{equation}\label{eq:4}
\frac{1}{\gamma_e^2}\left[\frac{\partial^2}{\partial t^2}+ 2\gamma_e\frac{\partial}{\partial t}+\gamma_e^2 - v_e^2 \nabla^2 \right]\phi_e(\mathbf{r},t) = Q_e(\mathbf{r},t).
\end{equation}
The form (\ref{eq:4}) results when the range distribution of corticocortical axons is approximated as isotropic, and exponentially decaying at large distances compared to characteristic range $r_e$ \citep{Robinson1997}. The firing rate field $\phi_e(\mathbf{r},t)$ is damped at a rate $\gamma_e = v_e/r_e$, where $v_e \simeq$ 5--10 m s$^{-1}$ is the average propagation rate along pyramidal axons.
For the remaining neuronal populations we assume propagation effects to be negligible, since their local interactions are relatively short-range. This leads to a small $r_a$ and large $\gamma_a$, and the simplified equation $\phi_a(t)=Q_a(t)$.

\subsection{Linearized equations}\label{sec:linear}

Fixed-point firing rate fields $\phi_a^{(0)}$ for a constant, uniform input $\phi_n$ are found by setting all time and spatial derivatives in Eqs. (\ref{eq:dend1})--(\ref{eq:4}) to zero \citep{Rennie1999, Robinson2005}. In the absence of perturbations the system will approach a stable fixed-point solution. Taylor expansion to first order around the fixed-point outgoing firing rates $Q_a^{(0)}$ yields
\begin{eqnarray}
Q_a(\mathbf{r},t) = Q_a^{(0)} + Q_a^{(1)},\\
Q_a^{(1)}=\rho_a \left[V_a(\mathbf{r},t)-V_a^{(0)}\right] = \rho_a V_a^{(1)}\label{eq:linearized_Q},
\end{eqnarray}
where $V_a^{(0)}$ is the equilibrium potential, and $\rho_a$ is the slope of the sigmoid at the fixed point, given by 
\begin{eqnarray}
\rho_a &=& \frac{d Q_a(\mathbf{r},t)}{d V_a(\mathbf{r},t)}\Big{|}_{V_a^{(0)}} \label{eq:rho1} \\
&=& \frac{\phi_a^{(0)}}{\sigma'}\left(1-\frac{\phi_a^{(0)}}{Q_{a}^{\mathrm{max}}} \right).\label{eq:rho2}
\end{eqnarray}
The above quantities lead to a set of gains
\begin{equation}\label{eq:6}
G_{ab} = \rho_a N_{ab}s_{ab} \equiv \rho_a \nu_{ab},
\end{equation}
giving the change in firing rate $\phi_a$ per unit change in afferent firing rate $\phi_b$. For products of gains representing loops or sequences of connections we will use the short-hand notation 
\begin{equation}\label{eq:combined_gains}
G_{abc} = G_{ab}G_{bc},
\end{equation}
and similarly for products of more than two gains. This leads to the notations listed in Table~1.

\begin{table}[ht]\label{table:gain_notations}
\centering
\begin{tabular}{lll}
\hline 
Number & Loop & Gain \\
\hline
1&Direct loop & $G_{esp_1d_1e}$ \\
2&Indirect loop & $G_{esp_1\varsigma p_2d_2e}$ \\
3&Alternative indirect loop with GPe-GPi/SNr projection & $G_{esp_1p_2d_2e}$ \\
4&Hyperdirect loop & $G_{esp_1\varsigma e}$ \\
5&GPe-STN loop & $G_{p_2\varsigma p_2}$ \\
6&Direct thalamocortical loop & $G_{ese}$ \\
7&Indirect thalamocortical loop through TRN & $G_{esre}$ \\
8&Intrathalamic loop & $G_{srs}$ \\
9&Basal ganglia-thalamic loop through D1 cells & $G_{d_1sp_1d_1}$ \\
10&Basal ganglia-thalamic loop through D2 cells & $G_{d_2sp_1\varsigma p_2d_2}$ \\
11&Basal ganglia-thalamic loop through D2 with GPe-GPi/SNr projection & $G_{d_2sp_1p_2d_2}$ \\
\hline
\end{tabular}
\caption{Gains for the various loops in the model as shown in Fig.~1.}
\end{table}

The linearized equation (\ref{eq:linearized_Q}) is solved more easily by passing to Fourier space, transforming from spatial and temporal coordinates $\mathbf{r}$ and $t$ to $\mathbf{k}$ and $\omega$. If we denote deviations from the fixed-point firing rates by $\phi_a^{(1)}$, substitution of the form of $V_a(\mathbf{k}, \omega)$ obtained by Fourier transforming (\ref{eq:dend1}) yields
\begin{eqnarray}
Q_a^{(1)}(\mathbf{k},\omega) &=& \rho_a V_a^{(1)}(\mathbf{k},\omega)\\
 &=& \left(1-\frac{i\omega}{\alpha} \right)^{-1}\left(1-\frac{i\omega}{\beta} \right)^{-1}\sum_b G_{ab}\phi_b^{(1)}(\mathbf{k}, \omega) e^{i\omega \tau_{ab}}.\label{eq:after_Fourier}
\end{eqnarray}
Also, Fourier transformation of (\ref{eq:4}) yields
\begin{eqnarray}\label{eq:split}
Q_a^{(1)}(\mathbf{k},\omega) =\left\{
     \begin{array}{ll}
          \left[\left(1-\frac{i\omega}{\gamma_e}\right)^2 + \frac{k^2v_e^2}{\gamma_e^2}\right]\phi_e^{(1)}(\mathbf{k},\omega) & (\mathrm{cortical~excitatory~neurons})\\
          \phi_a^{(1)}(\mathbf{k},\omega) & (\mathrm{all~other~populations})
     \end{array}
\right.
\end{eqnarray}
Expanding Eqs. (\ref{eq:after_Fourier}) and (\ref{eq:split}) for each of the components $a = e,i,d_1,d_2,p_1,p_2,\varsigma,s,$ and $r$ leads to a set of coupled linear equations that can be written in matrix form,
\setlength\arraycolsep{0.7pt}
\begin{eqnarray}\label{full_transfer}
  \left(\begin{array}{c}D_e\phi_e^{(1)}\\\phi_i^{(1)}\\\phi_{d_1}^{(1)}\\\phi_{d_2}^{(1)}\\\phi_{p_1}^{(1)}\\\phi_{p_2}^{(1)}\\\phi_{\varsigma}^{(1)}\\\phi_s^{(1)}\\\phi_r^{(1)} \end{array}\right) &=& L\left(\begin{array}{ccccccccc} G_{ee}& G_{ei}&0&0&0&0&0& K_{es}&0\\G_{ee}&G_{ei}&0&0&0&0&0&K_{es}&0\\ K_{d_1e}&0&G_{d_1d_1}&0&0&0&0&K_{d_1s}&0\\K_{d2e}&0&0&G_{d_2d_2}&0&0&0&K_{d_2s}&0\\0&0&K_{p_1d_1}&0&0&K_{p_1p_2}&K_{p_1\varsigma}&0&0\\0&0&0&K_{p_2d_2}&0&G_{p_2p_2}&K_{p_2\varsigma}&0&0\\K_{\varsigma e}&0&0&0&0&K_{\varsigma p_2}&0&0&0\\K_{se}&0&0&0&K_{sp_1}&0&0&0&K_{sr}\\K_{re}&0&0&0&0&0&0&K_{rs}&0 \end{array}\right) \left(\begin{array}{c}\phi_e^{(1)} \\ \phi_i^{(1)} \\ \phi_{d_1}^{(1)}\\\phi_{d_2}^{(1)}\\\phi_{p_1}^{(1)}\\\phi_{p_2}^{(1)}\\\phi_{\varsigma}^{(1)}\\\phi_{s}^{(1)}\\\phi_r^{(1)} \end{array}\right) + L\left(\begin{array}{c}0\\0\\0\\0\\0\\0\\0\\G_{sn}\phi_{n}^{(1)}\\0 \end{array}\right),\hspace{1cm}\label{eq:transfer_matrix}\\
D_e &=& \left(1-\frac{i\omega}{\gamma_e}\right)^2 + \frac{k^2v_e^2}{\gamma_e^2},\\
L &=& \left(1-\frac{i\omega}{\alpha} \right)^{-1}\left(1-\frac{i\omega}{\beta} \right)^{-1},\\
K_{ab} &=& G_{ab}e^{i\omega\tau_{ab}}.
\end{eqnarray}
%
%
\noindent Note that the differential operator $D_{\alpha\beta}$ associated with synaptodendritic transmission is transformed to $L$ in the Fourier domain, and differs from the operator $D_e$ associated with cortical wave propagation. Equation (\ref{eq:transfer_matrix}) allows us to derive a transfer function for input to the thalamus $\phi_n^{(1)}$ to the cortical excitatory firing rate $\phi_e^{(1)}$,
\begin{align}\label{transfer}
\frac{\phi_e^{(1)}(\mathbf{k}, \omega)}{\phi_{n}^{(1)}(\mathbf{k}, \omega)} =&  \frac{1}{k^2 r_e^2 + q^2 r_e^2} \frac{L^2e^{i\omega\tau_{es}}G_{es}G_{sn}}{(1-G_{ei}L)(1-T_{srs})M},\\
q^2 r_e^2 =& \left(1-\frac{i\omega}{\gamma_e}\right)^2 -  \frac{1}{1-G_{ei}L} \Bigg[G_{ee}L + \frac{T_{ese}+T_{esre}}{(1-T_{srs})M}\label{eq:dispersion_relation}\\\nonumber
+\frac{T_{esp_1d_1e}}{(1-T_{srs})(1-G_{d_1d_1}L)M}&+ \frac{T_{esp_1\varsigma e}}{(1-T_{srs})J~M} + \frac{T_{esp_1\varsigma p_2d_2e}+T_{esp_1p_2d_2e}}{(1-T_{srs})(1-G_{d_2d_2}L)(1-G_{p_2p_2}L)J~M}\Bigg], \hspace{1cm}\\
T_{a_1 a_2\ldots a_n} =& L^{n-1} G_{a_1 a_2}\ldots G_{a_{n-1}a_n}e^{i\omega(\tau_{a_1 a_2} + \ldots + \tau_{a_{n-1}a_n})}, \\
M = 1-&\frac{T_{d_1sp_1d_1}}{(1-T_{srs})(1-G_{d_1d_1}L)} - \frac{T_{d_2sp_1\varsigma p_2d_2}+T_{d_2sp_1p_2d_2}}{(1-T_{srs})(1-G_{d_2d_2}L)(1-G_{p_2p_2}L)J},\hspace{2cm}\\
J =& 1-\frac{T_{p_2\varsigma p_2}}{1-G_{p_2p_2}L}.
\end{align}
The transfer function is analogous to those customarily derived in control theory for linear systems \citep{Dorf2001}. Each of the $T_{a_1 a_2\ldots a_n}$ consists of a product of dendritic and synaptic filter functions $L(\omega)$, gains $G_{ab}$, and phase factor $e^{i\omega\tau_{ab}}$, representing one of the loops of the model: $T_{ese}$ and $T_{esre}$ are the two possible corticothalamic loops, the first from cortex to relay nuclei and back to cortex, and the second passing through the TRN, while $T_{esp_1d_1e}$ represents the classic direct pathway. Activity traveling along the indirect route, $T_{esp_1\varsigma p_2d_2e}$, the alternative indirect route, $T_{esp_1p_2d_2e}$, and the hyperdirect route, $T_{esp_1\varsigma e}$, is modulated by the loop between STN and GPe, $T_{p_2\varsigma p_2}$. This is expressed in the above equations by division of the relevant circuit strengths $T_{a_1a_2\ldots a_n}$ by $J$, which itself contains a factor to account for intrapallidal connections. The strengths of all pathways passing through the thalamus are modulated by the intrathalamic loop via $T_{srs}$. Pathways through the striatum have a correction factor that accounts for intrastriatal inhibition. Finally, all major pathways are modulated by the loops involving thalamostriatal projections, for which the correction factor is denoted $M$. The transfer function is considerably simplified if the thalamostriatal projection is removed, so that $M = 1$.

The electroencephalographic (EEG) signal is caused by extracellular currents near the dendrites of cortical neurons firing in partial synchrony \citep{Nunez1995, Ray1990}. As in previous works [e.g., \cite{Robinson2001, Robinson2003e}], we take the cortical potential to be proportional to the excitatory firing rate field $\phi_e$, since pyramidal neurons are the most aligned, most numerous, and largest cells contributing to the cortical signal. Hence, we approximate EEG spectral power as being proportional to $|\phi_e(\mathbf{k},\omega)|^2$. Parameter values in the normal state are given in Paper I, where their derivation and consistency with physiology are discussed extensively. As in that paper, we impose the random connectivity approximation \citep{Braitenberg1998, Robinson2001,Wright1995}, which leads to $\nu_{eb}=\nu_{ib}$ for $b = e,i,s$.

\subsection{Parameter changes in PD}\label{sec:changes}

In Paper I we considered five types of parameter changes for modeling the dopamine depletion that occurs in PD. These parameter changes are:\\
(I) decreasing both corticostriatal connection strengths and striatal firing thresholds to approximate a reduction in signal-to-noise ratio (SNR) in the striatum \citep{Leblois2006,Nicola2004, O'Donnell2003}, as modeled by
\begin{eqnarray}
\theta_{d_1}^{\mathrm{new}} = \theta_{d_1}-h\chi;\hspace{1cm} \theta_{d_2}^{\mathrm{new}} = \theta_{d_2}-h\chi,\label{eq:SNR1} \\
\nu_{d_1e}^{\mathrm{new}} = \nu_{d_1e} - \chi;\hspace{1cm} \nu_{d_2e}^{\mathrm{new}} = \nu_{d_2e} - \chi,\label{eq:SNR2}
\end{eqnarray} 
where $\chi \in [0,0.6]$ mV s, and $h = 5,10,15$ s$^{-1}$; \\
(II) increasing corticostriatal transmission to D2-expressing cells and decreasing transmission to D1-expressing cells, in line with the direct/indirect loop hypothesis \citep{Albin1989,Alexander1990,Mallet2006}; \\
(III) weakened lateral inhibition in the GPe due to enhanced release of enkephalin \citep{Stanford1999, Terman2002}; \\
(IV) weaker intracortical excitation and especially inhibition to approximate the effects of mesocortical dopamine loss \citep{Gulledge2001, Gao2003, Thurley2008,Zhou1999}; \\
(V) a combination of stronger cortico-D2 and weaker cortico-D1 transmission, weaker GPe-GPe inhibition, smaller cortical connection strengths, reduced GPe and STN firing thresholds, and a stronger D2-GPe projection. In particular, we considered the state that is obtained from the normal parameters in Paper I by setting $\nu_{d_1e} = 0.5$ mV s, $\nu_{d_2e} = 1.4$ mV s, $\nu_{p_2p_2}=-0.07$ mV s, $\nu_{ee}=\nu_{ie}=1.4$ mV s, $\nu_{ei}=\nu_{ii}=-1.6$ mV s, $\nu_{p_2d_2}=-0.5$ mV s, $\theta_{p_2} = 8$ mV, and $\theta_{\varsigma}=9$ mV. We will refer to this combination of parameters as the `full parkinsonian state'.

In this paper we consider the dynamical implications of dopamine depletion via the parameter changes (I)--(V), including responses to transient and ongoing stimuli, oscillations, and frequency spectra.

\section{Gains and responses to transient and ongoing stimuli}\label{sec:responses}

The linear gains (\ref{eq:6}) quantify the effective interactions between populations. In Sec~\ref{sec:gains} we consider changes in gains with nigrostriatal denervation, and propose a mechanism by which nigrostriatal and mesocortical dopamine loss can contribute to bradykinesia. Neuronal responses to transient cortical stimuli are derived in Sec. \ref{sec:transient_responses} using the full nonlinear version of the model. In Sec. \ref{sec:ongoing_responses} neuronal responses for ongoing brainstem inputs to the thalamus are derived.

\subsection{Changes in gains and a possible contribution to bradykinesia}\label{sec:gains}

 Linear gains depend on the connection strengths $\nu_{ab}$ not only directly, but also indirectly via the steady-state firing rates. The gains quantify the dynamic effects of changes in connection strengths. In the following, we discuss changes in gains obtained by manipulating the values of $\nu_{ab}$ to simulate dopamine loss, and recomputing steady states after each manipulation. The derivative of the sigmoid at each steady state is given by Eqs. (\ref{eq:rho1}) and (\ref{eq:rho2}), which yields gains via Eqs. (\ref{eq:6}) and (\ref{eq:combined_gains}). In the absence of thalamic inputs, the rate and amplitude of changes in cortical activity depend mainly on the sum $G_{ee}+G_{ei}$, which may be relevant to the symptoms of akinesia and bradykinesia, as further discussed below. Therefore, we report the value of this sum in each scenario, besides intracortical gains and the loop gains specified in Table~1.

\begin{figure}[!ht]
\centering
\includegraphics[width=420pt]{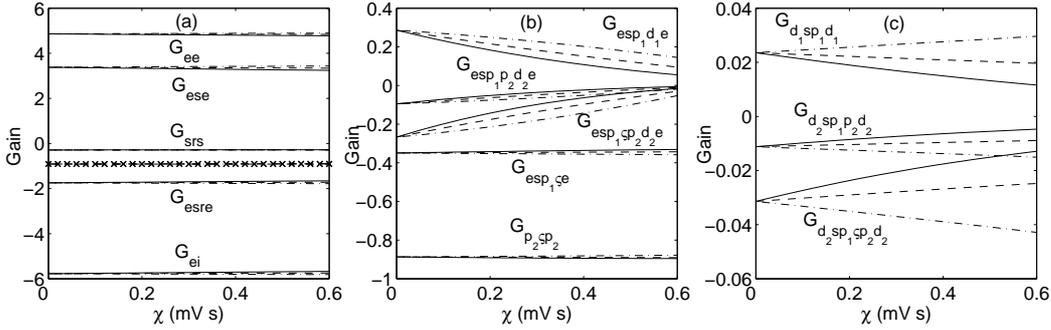}
\caption{Dependence of gains on the striatal SNR, parameterized by $\chi$. Solid lines, $h=5$ s$^{-1}$; dashed lines, $h=10$ s$^{-1}$; dash-dotted lines, $h=15$ s$^{-1}$ [cf. Eqs. (\ref{eq:SNR1}) and (\ref{eq:SNR2})]. Each group of lines starts together at $\chi=0$ mV s. (a) Corticothalamic gains. The lines of $G_{ee}+G_{ei}$ are indicated by the crosses, and overlap for different values of $h$. (b) Basal ganglia-thalamocortical gains (Loops 1--5 in Table~1). (c) Basal ganglia-thalamic gains (Loops 9--11 in Table~1).}
\label{fig:gains_vs_SNR}
\end{figure}

Simultaneous reduction of corticostriatal connection strengths and striatal firing thresholds has relatively little effect on gains, including the sum $G_{ee}+G_{ei}$ (cf. Fig. \ref{fig:gains_vs_SNR}). Especially corticothalamic gains are constant [Fig. \ref{fig:gains_vs_SNR}(a)], whereas both the direct and indirect loops become somewhat weaker [Fig. \ref{fig:gains_vs_SNR}(b)], and basal ganglia-thalamic gains may become stronger or weaker depending on $h$ [Fig. \ref{fig:gains_vs_SNR}(c); cf. Eq. (\ref{eq:SNR1})]. 

%

\begin{figure}[!ht]
\centering
\includegraphics[width=420pt]{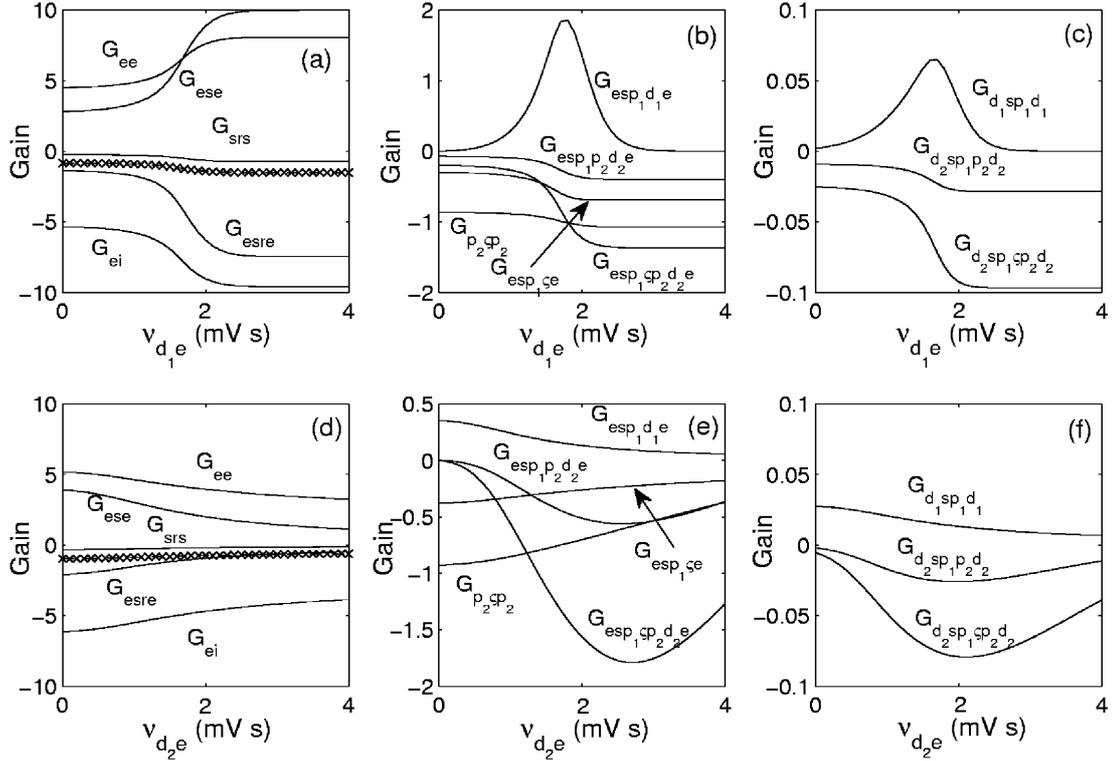}
\caption{Dependence of gains on corticostriatal connection strengths. (a) Corticothalamic gains vs. $\nu_{d_1e}$, with $G_{ee}+G_{ei}$ indicated by the crosses. (b) Gains of basal ganglia-thalamocortical loops vs. $\nu_{d_1e}$. (c) Gains of basal ganglia-thalamic loops vs. $\nu_{d_1e}$. (d) Corticothalamic gains vs. $\nu_{d_2e}$, with $G_{ee}+G_{ei}$ indicated by the crosses. (e) Gains of basal ganglia-thalamocortical loops vs. $\nu_{d_2e}$. (f) Gains of basal ganglia-thalamic loops vs. $\nu_{d_2e}$.}
\label{fig:gains}
\end{figure}

Figures \ref{fig:gains}(a) and \ref{fig:gains}(d) show that both increases in $\nu_{d_2e}$ and decreases in $\nu_{d_1e}$ reduce corticothalamic gains in absolute value. A slight reduction is also seen in $|G_{ee} + G_{ei}|$. A larger $\nu_{d_2e}$ enhances the strength of the indirect pathway while decreasing the strength of the direct pathway [cf. Figs. \ref{fig:gains}(b) and \ref{fig:gains}(e)]. On the other hand, decreasing $\nu_{d_1e}$ from its normal value of $1.0$ mV s weakens both the direct and indirect pathways. These changes in corticostriatal connection strengths decrease the absolute values of the gains of the STN-GPe loop, $G_{p_2\varsigma p_2}$, and the hyperdirect loop, $G_{esp_1\varsigma e}$. 

\begin{figure}[!ht]
\centering
\includegraphics[width=420pt]{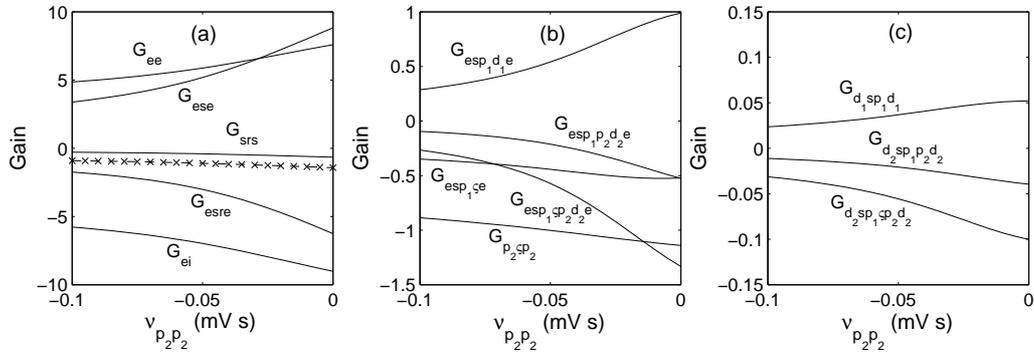}
\caption{Dependence of gains on the strength of lateral inhibition in the GPe. (a) Corticothalamic gains, with $G_{ee}+G_{ei}$ indicated by the crosses. (b) Basal ganglia-thalamocortical gains. (c) Basal ganglia-thalamic gains.}
\label{fig:gains_vs_nup2p2}
\end{figure}

The enhancement of corticothalamic gains seen in Fig. \ref{fig:gains_vs_nup2p2}(a) supports the view that reduced intrapallidal inhibition acts as a compensatory mechanism. However, Fig. \ref{fig:gains_vs_nup2p2}(b) reveals that all basal ganglia-thalamocortical circuits are strengthened by this change, including the direct, indirect, and hyperdirect loops, as well as the circuit formed by the STN and GPe. Thus, weakened intra-GPe inhibition may help stabilize basal ganglia firing rates, but exacerbate oscillations (cf. Sec. \ref{sec:full_stability}). 

As discussed in Paper I, a loss of direct dopaminergic inputs to the cortex is expected to reduce $\nu_{ee}, \nu_{ie}$, and especially $|\nu_{ei}|$ and $|\nu_{ii}|$. Since intracortical inhibition was taken to be stronger than excitation in the healthy state, this results in more similar strengths of excitation and inhibition, and a consequent decrease in $|G_{ee} + G_{ei}|$.

Reducing the GPe firing threshold increases the absolute values of corticothalamic gains, including $|G_{ee}+G_{ei}|$, as well as those of the direct, indirect, and hyperdirect loops, whereas lowering the STN firing threshold has the opposite effect. Reducing either $\theta_{\varsigma}$ or $\theta_{p_2}$ increases the gain of the STN-GPe loop. The indirect pathway is strengthened by greater $|\nu_{p_2d_2}|$, which weakens the direct, hyperdirect, and GPe-STN loops. Combining all parameter changes leads to a smaller $|G_{ee} + G_{ei}|$ (0.59 vs. 0.91 in the healthy state), a weaker direct loop ($G_{esp_1d_1e}=$ 0.042 vs. 0.29), and stronger hyperdirect ($G_{esp_1\varsigma e}=-0.40$ vs. $-0.35$), STN-GPe ($G_{p_2\varsigma p_2}=-1.1$ vs. $-0.89$), and especially indirect ($G_{esp_1\varsigma p_2d_2e}=-3.0$ vs. $-0.27$) loops in the parkinsonian state compared to the healthy state.

\begin{figure}[!ht]
\centering
\includegraphics[width=420pt]{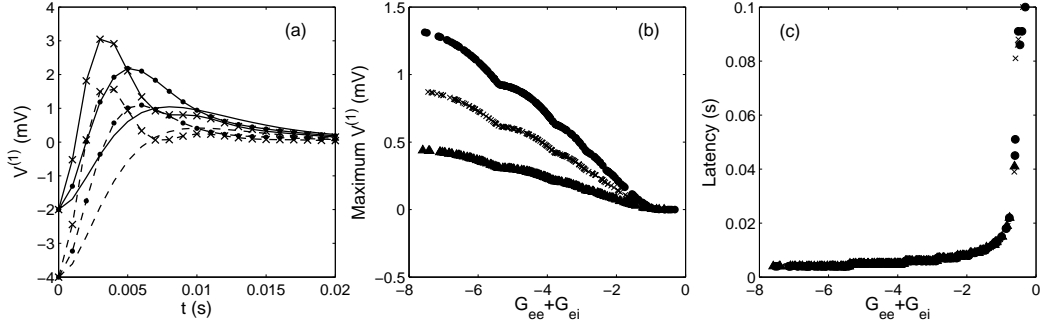}
\caption{Dependence of cortical response strength and latency on gains. Time derivatives have initial values $dV_e^{(1)}/dt=dV_i^{(1)}/dt=0$ mV s$^{-1}$ in all cases. (a) Time courses of $V_e^{(1)}$ (solid) and $V_i^{(1)}$ (dashed), with $V_e^{(1)}(0) = -2$ mV and $V_i^{(1)}(0) = -4$ mV, for different values of the intracortical gains. Crosses, $G_{ee} = 3, G_{ei}=-11$; dots, $G_{ee} = 2, G_{ei}=-5$; no markers, $G_{ee}=1, G_{ei}=-2$. (b) Maximum values of $V_e^{(1)}$ and $V_i^{(1)}$ vs. $G_{ee}+G_{ei}$. Dots, $V_e^{(1)}(0) = V_i^{(1)}(0) = -6$ mV; crosses, $V_e^{(1)}(0) = V_i^{(1)}(0) = -4$ mV; triangles, $V_e^{(1)}(0) = V_i^{(1)}(0) = -2$ mV. (c) Latency of the maximum vs. $G_{ee}+G_{ei}$.}
\label{fig:control}
\end{figure}

We investigate the results of changes in cortical gains by considering only cortical interactions and ignoring the damped-wave equation (\ref{eq:4}), which leads to $\phi_a=Q_a$. Substituting $a=e,i$ in Eqs. (\ref{eq:dend1}) and (\ref{eq:dend2}), and using the linear approximation $Q_a=V_a^{(0)}+\rho_a V_a^{(1)}$ yields
\begin{eqnarray}
\frac{d^2 V_e^{(1)}}{dt^2}=\alpha \beta\left[G_{ee}V_e^{(1)} + G_{ei}V_i^{(1)} - \left(\frac{1}{\alpha} + \frac{1}{\beta}\right)\frac{dV_e^{(1)}}{dt}-V_e^{(1)}\right],\label{eq:num1}\\
\frac{d^2 V_i^{(1)}}{dt^2}=\alpha \beta\left[G_{ee}V_e^{(1)} + G_{ei}V_i^{(1)} - \left(\frac{1}{\alpha} + \frac{1}{\beta}\right)\frac{dV_i^{(1)}}{dt}-V_i^{(1)}\right].\label{eq:num2}
\end{eqnarray}
Here, we have made use of the random connectivity approximation to obtain $G_{ie}=G_{ee}$ and $G_{ii}=G_{ei}$. The sizes and latencies of the maximums of $V_e^{(1)}$ and $V_i^{(1)}$ were determined using numerical integration of Eqs. (\ref{eq:num1}) and (\ref{eq:num2}) for three sets of initial conditions, using 200 pairs of values of $G_{ee}$ and $G_{ei}$ uniformly distributed in the intervals $[1,5]$ and $[-9,-5]$. As illustrated in Fig. \ref{fig:control}, response strength and latency do not depend on $G_{ee}$ or $G_{ei}$ individually, but input responses become slower and less pronounced with decreased $|G_{ee}+G_{ei}|$. The decreased value of $|G_{ee}+G_{ei}|$ due to increased strength of the indirect pathway and/or impaired cortical inhibition after loss of mesocortical dopamine may thus be related to slowness of movement (bradykinesia) or absence of movement (akinesia), since threshold activation levels necessary to initiate movements may be reached more slowly, or not at all. This hypothesis is supported by the observation that MPTP lesion in monkeys causes the activity of motor cortical neurons to build up more slowly and persist longer during voluntary movements \citep{Doudet1990}. Similarly, transcranial magnetic stimulation leads to more gradual modification of motor unit activity in PD patients than in controls \citep{Kleine2001}.

\subsection{Responses to transient stimuli}\label{sec:transient_responses}

\begin{figure}[!ht]
\centering
\includegraphics[width=420pt]{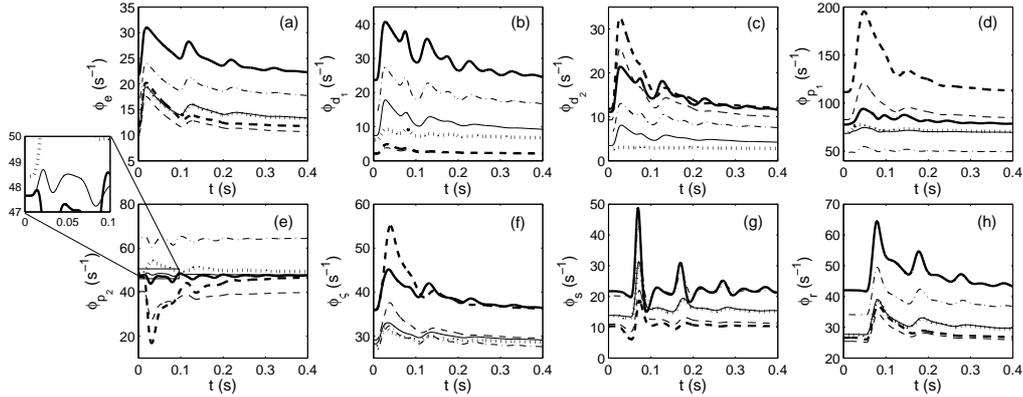}
\caption{Responses of firing rate fields to a square pulse of 10 ms duration and amplitude 60 s$^{-1}$ applied at the cortex, while the relay nuclei receive a constant brainstem input $\phi_n=10$ s$^{-1}$. Thin solid lines, healthy state with parameters as in Paper I; dotted, `reduced-SNR' state with $\theta_{d_1}=\theta_{d_2}=13$ mV, $\nu_{d_1e} = 0.4$ mV s, and $\nu_{d_2e} = 0.1$ mV s; thin dashed lines, state with $\nu_{d_1e} = 0.5$ mV s, $\nu_{d_2e} = 1.4$ mV s; dash-dotted, with weaker intrapallidal inhibition, $\nu_{p_2p_2} = -0.03$ mV s; thick solid lines, healthy state with $\nu_{ee}=\nu_{ie}=1.4$ mV s and $\nu_{ei}=\nu_{ii}=-1.6$ mV s to model cortical dopamine loss; thick dashed lines, full parkinsonian state (cf. Sec. \ref{sec:changes}). The inset to (e) shows the triphasic early GPe response.}
\label{fig:impulse_responses}
\end{figure}

Figure \ref{fig:impulse_responses} shows the responses in each neural population to a 10 ms square pulse with amplitude 60 s$^{-1}$ applied at the cortex in the healthy condition with parameters as in Paper I, and with the five possible results of dopamine loss mentioned in Sec. \ref{sec:changes}. To compare these responses to experimental results, we make use of the finding that the average firing rate of motor cortical neurons is increased during movement \citep{Grammont2003,Thach1978}.

In the healthy condition, the slight increase in the model rate of the output nuclei with cortical stimulation is in agreement with experimental findings that the majority of GPi neurons increase their activity before and during movement \citep{Anderson1985,Georgopoulos1983, Mitchell1987}, presumably because of inhibition of competing motor programs when a target program is activated \citep{Nambu2002}. The latencies and directions of GPe responses seen in the insert to Fig. \ref{fig:impulse_responses}(e) closely match those observed upon stimulation of M1, SMA, or S1 in healthy awake monkeys: an early excitation after 8--11 ms, inhibition after 15--19 ms, and a late excitation after 26--32 ms \citep{Kita2004, Nambu2000, Yoshida1993}. With the parameters in Paper I, onset of GPe excitation by the STN occurs 8 ms after cortical stimulation, inhibition by striatum after 19 ms, and the second excitation by STN after 28 ms. The relative sizes of the peaks and trough depend on the stimulation intensity, longer or faster stimuli leading to a deeper trough. However, excitations are stronger than inhibition for 10 ms inputs up to at least $60$ s$^{-1}$, in accord with predominantly excitatory responses to movement in healthy monkeys \citep{Anderson1985,Mitchell1987, Turner1997}. The model and physiological data suggest that this excitation is mediated mainly via the hyperdirect pathway. In reality, movements may elicit stronger GPe excitation in view of the close association of cortico-STN fibers with the pyramidal tract \citep{Giuffrida1985,Nambu2002}, but we do not distinguish between cortical neurons projecting to striatum or STN in our model. In healthy African green monkeys, \citet{Bergman1994} found that the majority of STN cells increased their firing rates when torque was applied to the elbow, although cells that decreased their firing rates did so for longer. It is seen in Fig. \ref{fig:impulse_responses}(f) that the model STN rate displays a transient increase upon cortical stimulation, in agreement with these results. 

Considering now the parkinsonian scenarios, reducing the SNR has little effect on most responses to a transient stimulus, and cortical and thalamic responses are virtually indistinguishable from normal ones (cf. Fig. \ref{fig:impulse_responses}). Striatal inhibition of GPe falls away, resulting in an amplified excitatory response to STN input. Corticothalamic responses are slightly attenuated when modeling dopamine loss with $\nu_{d_1e}=0.5$ mV s and $\nu_{d_2e} = 1.4$ mV s, as expected from the smaller gains. These gain changes were inferred from the increased responsiveness of D2 cells and decreased responsiveness of D1 cells to cortical inputs measured experimentally, and are thus in accord with these results \citep{Cepeda1998,Hernandez-Lopez1997,Hsu1995,Levine1996, Kiyatkin1996,Nicola2000, Toan1985, Umemiya1997}. The larger increase in the firing rate of the output nuclei seen in Fig. \ref{fig:impulse_responses}(d) is in line with amplified GPi responses to passive limb movements in  parkinsonian monkeys \citep{Bergman1994,Filion1988,Wichmann1994}. The inhibitory response in the GPe is greatly amplified in this scenario, as seen in Fig. \ref{fig:impulse_responses}(e). This reproduces the effects of extensive nigrostriatal lesions in rats \citep{Magill2001, Zold2007}, although facilitated excitation has also been observed \citep{Tremblay1989}. Finally, the more vigorous responses of STN neurons to cortical stimulation in parkinsonian rats, and to elbow flexion and extension in monkeys, corroborates the results in Fig. \ref{fig:impulse_responses}(f) \citep{Bergman1994,Magill2001}. Figures \ref{fig:impulse_responses}(a) shows that weakened intrapallidal inhibition ($\nu_{p_2p_2}=-0.03$ mV s) amplifies damped oscillations around the dominant alpha frequency. An even stronger effect on corticothalamic interactions is exerted by smaller cortical gains, causing an amplification of damped oscillations at alpha and beta frequencies throughout the system, as seen in Figs. \ref{fig:impulse_responses}(a) and  \ref{fig:impulse_responses}(g). Damped corticothalamic oscillations are enhanced very slightly by a lower GPe firing threshold, whereas a lower STN threshold and larger $|\nu_{p_2d_2}|$ have the opposite effect. The full parkinsonian state produces changes in responses similar to those resulting only from stronger indirect and weaker direct pathways, and with a cortical rate that is closer to that in the healthy state, in line with experimental observations. 

\subsection{Responses to ongoing stimuli}\label{sec:ongoing_responses}

\begin{figure}[!hp]
\centering
  \includegraphics[width=420pt]{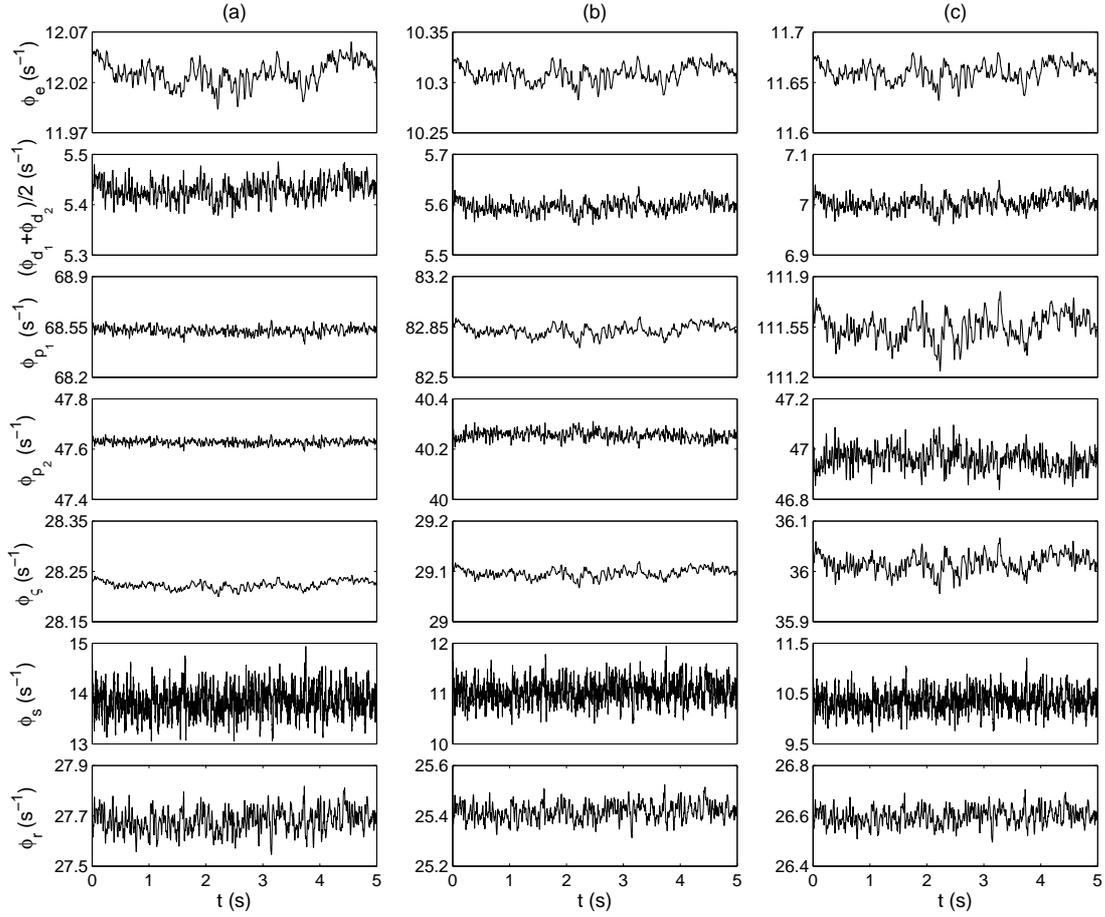}
  \caption{Time series of the cortical ($\phi_e$), striatal ($[\phi_{d_1}+\phi_{d_2}]/2$), GPi/SNr ($\phi_{p_1}$), GPe ($\phi_{p_2}$), STN ($\phi_{\varsigma}$), thalamic relay ($\phi_s$), and TRN ($\phi_r$) firing rate fields (a) in the normal state, (b) with $\nu_{d_1e}=0.5$ mV s and $\nu_{d_2e}=1.4$ mV s, and (c) in the full parkinsonian state (cf. Sec. \ref{sec:changes}). Input to the relay nuclei consisted of Gaussian white noise with mean 10 s$^{-1}$ and standard deviation 2 s$^{-1}$.} 
\label{fig:detailed_timeseries}
\end{figure}

Figure \ref{fig:detailed_timeseries} shows pulse rates $\phi_a$ for ongoing stimuli $\phi_n$ in the healthy and two model parkinsonian-type states. In each case the stimulus consisted of Gaussian noise with mean 10 s$^{-1}$ and standard deviation 2 s$^{-1}$. Firing rates of each component are plotted over equal intervals along the ordinate to allow comparison of variability across states. Figure \ref{fig:detailed_timeseries} reveals relatively large fluctuations in pallidal and STN firing rates in modeled parkinsonian states,
even in the GPe, where the average firing rate is lower than in the healthy state. These enhanced fluctuations point towards increased synchronization between individual cells in the basal ganglia nuclei. On the other hand, cortical, thalamic, and average striatal signals show reduced variability. The amplitude of fluctuations in the average signal of D1 and D2 cells depends on the relative changes in $\nu_{d_1e}$ and $\nu_{d_2e}$, and is elevated compared to the healthy state for relatively large increases in $\nu_{d_2e}$. Mesocortical dopamine loss and changes secondary to nigrostriatal damage normalize the amplitude of the cortical signal, and further amplify basal ganglia fluctuations, as seen in Fig. \ref{fig:detailed_timeseries}(c). In Sec. \ref{sec:full_spectra} we will relate these observations to changes in frequency spectra.

\section{Oscillations and spectral changes}\label{sec:oscillations}

In certain regimes the model displays oscillations that may culminate in limit cycles. Section~\ref{sec:full_stability} explores these oscillations, and in particular, we find approximately 5 Hz oscillations in the indirect loop and $\sim$20 Hz oscillations in corticothalamic circuits that may spread to the basal ganglia when the indirect pathway becomes strong. The spectral changes caused by dopamine loss are considered in Sec.~\ref{sec:full_spectra}. The involvement of the indirect pathway in the generation of parkinsonian symptoms has been questioned \citep{Leblois2006} because GPe lesion does not lead to the characteristic motor symptoms or oscillations \citep{Soares2004}. In Sec.~\ref{sec:GPe_lesion} we challenge the view that GPe lesion experiments exclude the indirect pathway as a substrate of parkinsonian oscillations.

\subsection{Limit cycles and the emergence of theta and beta rhythms}\label{sec:full_stability}

The linearized equations given in Sec. \ref{sec:linear} are only valid in a limited regime. Far from a steady state, the system may be attracted to a different fixed point or a limit cycle. Such deviations from steady state entail changes in gains, and we can vary the gains to determine the boundaries of the linearly stable region. A relationship between the frequencies of wave modes and their wavelengths is termed a dispersion relation. The transfer function (\ref{transfer}) has an associated dispersion relation $k^2 + q^2(\omega) = 0$. For each combination of gain values, the dispersion relation has a specific (usually infinite) set of solutions $\omega(\mathbf{k})$. Of the (complex) solutions $\omega(\mathbf{k})$, the one with the largest imaginary part decays most slowly, since solutions consist of a weighted sum of plane waves $e^{-i\omega(k) t + i\mathbf{k}\mathbf{r}}$. The boundary of the linearly stable zone occurs where the dispersion relation is satisfied for real $\omega$, because for Im $\omega > 0$ inputs $\phi_n^{(1)}$ are infinitely amplified at the corresponding frequencies. Previous work has shown that the spatially uniform mode is generally the least stable \citep{Robinson2001b, Robinson2005b}, leading us to consider this $\mathbf{k} = \mathbf{0}$ case. Due to the complexity of our system, instabilities lead to resonances at a range of frequencies depending on the gain being varied. \\

\noindent \emph{Perturbations of healthy state}

We first consider the types of instabilities that can arise from perturbations around the healthy state, which is the low-firing rate fixed point corresponding to the parameters in Paper I. Table~2 lists the frequencies of instabilities for changes in gains relative to the healthy state. The gains at which the instabilities first occur are also given. For completeness, limits are listed for both increases and decreases in gains, partly since the signs of effective interactions are not known with certainty in every case (this applies especially to intrastriatal interactions). Some instabilities will not readily occur because of an extremely large threshold value, for instance for $G_{p_2d_2}$ and $G_{p_1\varsigma}$. At a 0 Hz instability, the system may shift to a different steady state or go into a limit cycle if one exists. 

\definecolor{gray}{rgb}{0.7,0.7,0.7}

\begin{table}[htp]\label{instab_freq}
\centering
\begin{tabular}{p{1.7cm} p{1cm} p{2.7cm} p{1.7cm} p{2.7cm} p{1.7cm}}
\hline
Gain & Sign & Upper threshold & $f$ (Hz) & Lower threshold & $f$ (Hz) \\ 
\hline
$G_{ee}$, $G_{ie}$ & $+$ & 5.6 & \colorbox {gray}{0} & --- & --- \\
$G_{ei}$, $G_{ii}$ & $-$ &  $-5.0$ & \colorbox {gray}{0} & --- & --- \\
$G_{es}$, $G_{is}$ & $+$ & 2.0 & \colorbox {gray}{0} & $-3.1$ & 4.1 \\
$G_{d_1e}$ & $+$ & 10 & \colorbox {gray}{0} & $-59$ & 5.8 \\
$G_{d_1d_1}$ & $-$ & 0.71 & 0 & --- & --- \\
$G_{d_1s}$ & $+$ & 5.6 & \colorbox {gray}{0} & $-17$ & 20 \\
$G_{d_2e}$ & $+$ & 41 & \colorbox {gray}{5.0} & $-5.2$ & 0 \\
$G_{d_2d_2}$ & $-$ & 1.0 & 1.3 & --- & --- \\
$G_{d_2s}$ & $+$ & 12 & \colorbox {gray}{17} & $-3.6$ & 0 \\
$G_{p_1d_1}$ & $-$ & 66 & 20 & $-6.9$ &\colorbox {gray}{0} \\
$G_{p_1p_2}$ & $-$ & 130 & 4.0 & $-61$ & \colorbox {gray}{0} \\
$G_{p_1\varsigma}$ & $+$ & 94 & \colorbox {gray}{6.0} & $-7.1$ & 0 \\
$G_{p_2d_2}$ & $-$ & 21 & 0 & $-260$ & \colorbox {gray}{18} \\
$G_{p_2p_2}$ & $-$ & 1.6 & 10 & --- & --- \\
$G_{p_2\varsigma}$ & $+$ & 21 & \colorbox {gray}{46} & $-7.0$ & 2.0 \\
$G_{\varsigma e}$ & $+$ & 23 & \colorbox {gray}{6.1} & $-2.5$ & 0 \\
$G_{\varsigma p_2}$ & $-$ & 0.66 & 0.42 & $-1.8$ & \colorbox {gray}{46} \\
$G_{se}$ & $+$ & 3.5 & \colorbox {gray}{0} & $-1.8$ & 3.7 \\
$G_{sp_1}$ & $-$ &  0.80 & 0 & $-5.2$ & \colorbox {gray}{5.7} \\
$G_{sr}$ & $-$ & $-0.82$ & \colorbox {gray}{0} & $-10$ & \colorbox {gray}{3.6} \\
$G_{re}$ & $+$ & 4.1 & \colorbox {gray}{3.3} & $0.48$ & \colorbox {gray}{0} \\
$G_{rs}$ & $+$ & 1.8 &\colorbox {gray}{30} & $-0.14$ & 0 \\
\hline
\end{tabular}
\caption{Frequencies $f$ of linear instabilities caused by increasing or decreasing individual gains beyond the given threshold values, corresponding to the boundary of the linearly stable region. All other gains are held constant at the values for the normal state, i.e., the low-firing rate steady state for the parameters given in Paper I. The first six gains are varied in pairs, because they are equal in the random connectivity approximation (cf. Sec. 3.2). No instabilities occur for reductions in $G_{d_1d_1}$, $G_{d_2d_2}$, $G_{p_2p_2}$, $G_{ee}$ and $G_{ie}$, or $G_{ei}$ and $G_{ii}$. The second column indicates whether the gain is expected to be excitatory ($+$) or inhibitory ($-$) based on physiological considerations. Frequencies of instabilities for which the gain has the expected sign are shaded. All values are given to two significant figures.}
\end{table}

Many of the instabilities in Table~2 are approximately equivalent, because the dispersion relation contains products and ratios of gains. This is clearest for $G_{p_2\varsigma}$ and $G_{\varsigma p_2}$, since either raising one or decreasing the other leads to an instability at 46 Hz. A resonant frequency in the gamma band for the STN-GPe loop corresponds with the $\sim$55 Hz oscillation in the model of \citet{Humphries2006}, and identifies this loop as a possible substrate for the enhanced gamma oscillations seen in parkinsonian patients on levodopa right before and during voluntary movements \citep{Brown2003, Cassidy2002}. 

Other gains that occur together in a loop cause instabilities at different frequencies, because they simultaneously modulate at least one other loop. For some of these gains, the loops that sustain the oscillations are relatively easy to determine. For instance, the gain $G_{sr}$ is part of the thalamocortical loop strength $G_{esre}$, whereas $G_{rs}$ is only relevant to the intrathalamic loop, $G_{srs}$. Since $G_{sr}$ and $G_{rs}$ cause instabilities at 3.6 Hz and 30 Hz, respectively, we can conclude that thalamocortical instabilities occur around 3--4 Hz, and intrathalamic rhythms have a frequency of about 30 Hz. In the absence of cortical and corticothalamic feedback and for $\tau_{sr}+\tau_{rs}=0$, this frequency becomes $f=\sqrt{\alpha\beta}/(2\pi)=51$ Hz, analogous to spindle instabilities described by \citet{Robinson2002}. The 3--4 Hz instability leads to a limit cycle at approximately the same frequency, which often has a spike-wave form and was shown by \citet{Robinson2002} and \citet{Breakspear2006} to have many of the characteristics of petit mal (or absence) seizures. 

Other interactions between loops are more difficult to untangle. For example, the gain $G_{d_2e}$ is part of the indirect and alternative indirect loops 2 and 3 (cf. Table~1 for loop numbering), while $G_{p_2d_2}$ is part of the same loops and also the basal ganglia-thalamic loops 10 and 11. The basal ganglia-thalamic circuits through D2 support oscillations around 17 Hz for large $G_{d_2s}$. The frequency of the instability due to a large negative $G_{p_2d_2}$ ($\sim$18 Hz) is inversely related to axonal delays in the indirect loop, but also to $\tau_{d_2s}$ and $\tau_{se}$. This suggests that the oscillations are sustained by a complex interplay between corticothalamic circuits and the indirect loop, rather than any particular circuit separately. The gain from the output nuclei to the thalamus, $G_{sp_1}$, is part of the hyperdirect, direct, classic and alternative indirect (with the GPe-GPi/SNr projection) pathways, as well as loops involving thalamostriatal projections (Loops 1--4 and 9--11). The frequency of the instability for large $G_{sp_1}$ is 5.7 Hz, and depends inversely on $\tau_{\varsigma e}, \tau_{p_1\varsigma}$, and $\tau_{es}$, with a weaker dependence on $\tau_{d_2e}, \tau_{p_2d_2}, \tau_{d_1s}, \tau_{d_2s}$, or $\tau_{p_1p_2}$. This indicates that the oscillations are sustained by the hyperdirect pathway. Although the frequency due to strong $G_{\varsigma e}$ (6.1 Hz) is close to that of tremor rhythms, we saw in Sec. \ref{sec:gains} that neither a decreased striatal SNR, nor a stronger indirect and weaker direct loop, lead to a large hyperdirect loop gain. This suggests that limit cycle oscillations are unlikely to arise in the hyperdirect loop, as argued further below.\\

\noindent \emph{Perturbations leading to a strong indirect pathway: theta oscillations}

\begin{figure}[!ht]
\centering
\includegraphics[width=420pt]{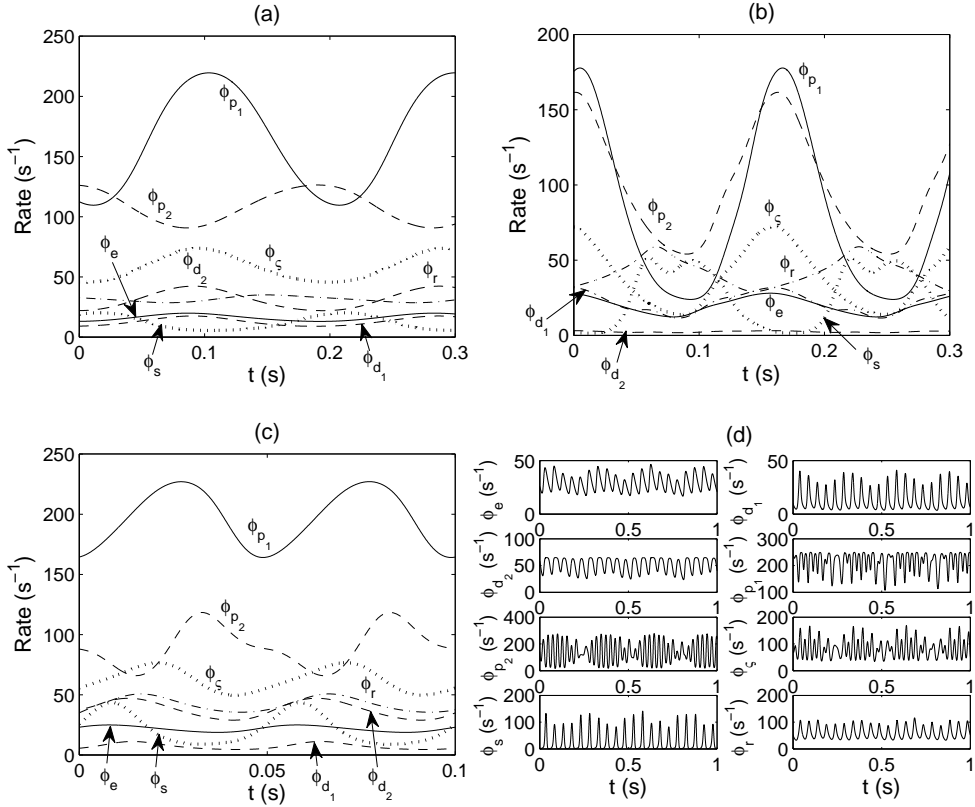}
\caption{Possible limit cycles in the BGTCS. (a) Limit cycle at 5.1 Hz caused by a moderate corticothalamic gain and a much larger indirect loop gain ($G_{esp_1\varsigma p_2d_2e}=-49$, $G_{ese}=2.7$). The GPe oscillates approximately in antiphase with the STN. (b) Limit cycle at 6.2 Hz caused by a strong hyperdirect loop ($G_{esp_1\varsigma e}=-52$). STN and GPe oscillate approximately in phase. (c) Limit cycle at 20 Hz caused by a large corticothalamic gain and a moderately larger indirect loop gain ($G_{esp_1\varsigma p_2d_2e}=-32$, $G_{ese}=7.9$). Note that the scale differs from (a) and (b). (d) Oscillations arising for a large corticothalamic gain and a much larger indirect loop gain ($G_{esp_1\varsigma p_2d_2e}=-76$, $G_{ese}=8.1$). The strongest resonance occurs at 20 Hz, with a weaker resonance at 3.6 Hz.}
\label{fig:limit_cycles}
\end{figure}

Since electrophysiological evidence and modeling results on firing rates (cf. Paper I) suggest that dopamine depletion increases the gain of the indirect pathway, we look at oscillations that appear when the balance is shifted from the direct to the indirect pathway. For parameter values that entail a sufficiently large increase of the indirect loop gain, the system undergoes a Hopf bifurcation, leading to a limit cycle around 5 Hz. In Fig. \ref{fig:limit_cycles}(a) we consider $G_{esp_1\varsigma p_2d_2e}=-49$, which is instantiated by the parameters $\theta_{p_2}=4$ mV, $\theta_{\varsigma}=5$ mV, $\nu_{ee}=\nu_{ie}=1.2$ mV s, $\nu_{ei}=\nu_{ii}=-1.3$ mV s, $\nu_{d_2e}=1.7$ mV s, $\nu_{p_2d_2}=-0.5$ mV s, $\nu_{p_2p_2}=0$ mV s, $\nu_{sp_1}=-0.07$ mV s, $\nu_{se}=0.6$ mV s, and $\phi_n=30$ s$^{-1}$. Since the indirect loop has a negative overall gain, deviations from the mean activity are inverted after one pass around the loop, so that a single period of the oscillation corresponds to two passes, similar to oscillations in corticothalamic circuits involving the TRN \citep{Robinson2002}. Approximately 5 Hz oscillations only appear when the ratio of the indirect loop and corticothalamic gains is very large, although the latter still need to be sufficiently powerful for the system to support oscillations (here, $G_{ese} = 2.7$). Considering the extreme parameter values and firing rates, such limit cycle behavior is unlikely to occur in the system as a whole, but could appear in subcircuits, which would explain the limited percentage of oscillatory cells generally recorded \citep{Bergman1994,Lemstra1999,Levy2001, Wichmann2006}. This scenario is further supported by the finding that oscillatory cells in the striatum, STN, and GPi have substantially higher firing rates than non-oscillatory cells \citep{Dejean2008,Levy2000,Levy2001}. Significantly, phase relationships between components comply with those in rats with nigrostriatal lesions \citep{Walters2007}: GPe activity is in antiphase with striatal, STN, cortical, and SNr activity. The fact that the GPe oscillates antiphase to STN implies that it is under inhibitory control of the striatum rather than excitatory control of the STN. A caveat is that the oscillations reported by \citet{Walters2007} occurred around 1 Hz, and may have arisen partly as a result of urethane anesthesia \citep{Humphries2006,Magill2000}. However, urethane-induced oscillations in GP appear to be in phase with cortical activity in the healthy state, whereas nigrostriatal lesion produces a subpopulation oscillating in antiphase to cortical slow waves \citep{Zold2007b}. \\

\noindent \emph{Perturbations leading to a strong hyperdirect pathway}

Increasing the gain of the hyperdirect pathway (Loop 4) leads to phase relationships that are not supported by experiments. As an example, Fig. \ref{fig:limit_cycles}(b) shows the 6.2 Hz limit cycle corresponding to $G_{esp_1\varsigma e}=-52$, obtained by setting $\theta_{p_2}=4$ mV, $\theta_{\varsigma}=20$ mV, $\nu_{d_2e}=0.3$ mV s, $\nu_{\varsigma e}=0.7$ mV s, $\nu_{sp_1}=-0.15$ mV s, and $\phi_n=30$ s$^{-1}$. In this case, the GPe oscillates nearly in phase with the striatum, STN, cortex, and SNr, indicating that it is driven by STN input. \\

\noindent \emph{Perturbations leading to a strong indirect pathway: beta oscillations}

As discussed above in relation to increases in $G_{p_2d_2}$, it is also possible to obtain $\sim$20 Hz oscillations when the indirect loop is strong, as seen experimentally in PD patients \citep{Brown2001, Fogelson2006, Gatev2006, Levy2002, Williams2002}. This happens particularly in combination with corticothalamic coupling that is strong in comparison to that leading to 5 Hz oscillations, while the ratio $|G_{esp_1\varsigma p_2d_2e}/G_{ese}|$ can be somewhat smaller. An example with $G_{esp_1\varsigma p_2d_2e}=-32$ and $G_{ese}=7.9$ is presented in Fig. \ref{fig:limit_cycles}(c). The corresponding parameters are $\theta_{p_2}=7$ mV, $\theta_{\varsigma}=6$ mV, $\nu_{ee}=\nu_{ie}=0.6$ mV s, $\nu_{ei}=\nu_{ii}=-0.8$ mV s, $\nu_{d_1e}=0.5$ mV s, $\nu_{d_2e}=1.4$ mV s, $\nu_{p_2d_2}=-0.4$ mV s, $\nu_{p_2p_2}=0$ mV s, $\nu_{sp_1}=-0.04$ mV s, and $\phi_n=20$ s$^{-1}$. The oscillations first appear in the cortex and thalamus, from where they spread to the basal ganglia. Approximately 20 Hz oscillations can also arise in corticothalamic circuits in the healthy state, seen both experimentally \citep{Courtemanche2003} and in our model. It is possible that either the enhancement of these oscillations in the basal ganglia of PD patients contributes to antikinetic symptoms or tremor, or they may be secondary to other pathological activity rather than a cause of symptoms.\\

\noindent \emph{Perturbations leading to a strong indirect pathway: combined theta and beta}

Our model predicts that if beta oscillations appear in the corticothalamic loop in the absence of nigrostriatal damage, then strong alpha oscillations will also be seen, since beta rhythms arise as a harmonic of the alpha resonance \citep{Robinson2001}. However, $\sim$20 Hz and 3--7 Hz oscillations can appear together without $\sim$10 Hz activity when the alpha resonance is suppressed by nigrostriatal damage. Both $G_{ese}$ and the ratio $|G_{esp_1\varsigma p_2d_2e}/G_{ese}|$ need to be large for this situation to occur. An example is given in Fig. \ref{fig:limit_cycles}(d) for $G_{esp_1\varsigma p_2d_2e}=-76$ and $G_{ese}=8.1$, with parameter values $\theta_{p_2}=5$ mV, $\theta_{\varsigma}=4$ mV, $\nu_{ee}=\nu_{ie}=0.5$ mV s, $\nu_{ei}=\nu_{ii}=-0.7$ mV s, $\nu_{d_1e}=0.5$ mV s, $\nu_{d_2e}=1.4$ mV s, $\nu_{p_2d_2}=-0.4$ mV s, $\nu_{p_2p_2}=0$ mV s, $\nu_{sp_1}=-0.07$ mV s, and $\phi_n=30$ s$^{-1}$. The strong interaction between the theta and alpha-band roots of the dispersion relation relies on the relatively small distance between these roots, explaining why the indirect loop can suppress alpha, but not beta, activity. This is further discussed in Sec. \ref{sec:full_spectra}.\\

All the above limit cycles have in common that they require relatively strong corticothalamic activation to arise, provided for instance by brainstem input to the thalamus or reduced intracortical inhibition, potentially helping to explain the worsening of tremor during mental stress \citep{Deuschl2001, Zesiewicz2001}. The direct loop can also contribute to excitation of the cortex and thalamus, so that 5 Hz and 20 Hz oscillations can appear when both the direct and indirect loops are strong. This implies that the direct and indirect loops need not be completely separated for basal ganglia-thalamocortical loops to support theta and beta oscillations.

\begin{figure}[!ht]
\centering
\subfigure{\label{fig:tent}\includegraphics[width=220pt, height=160pt]{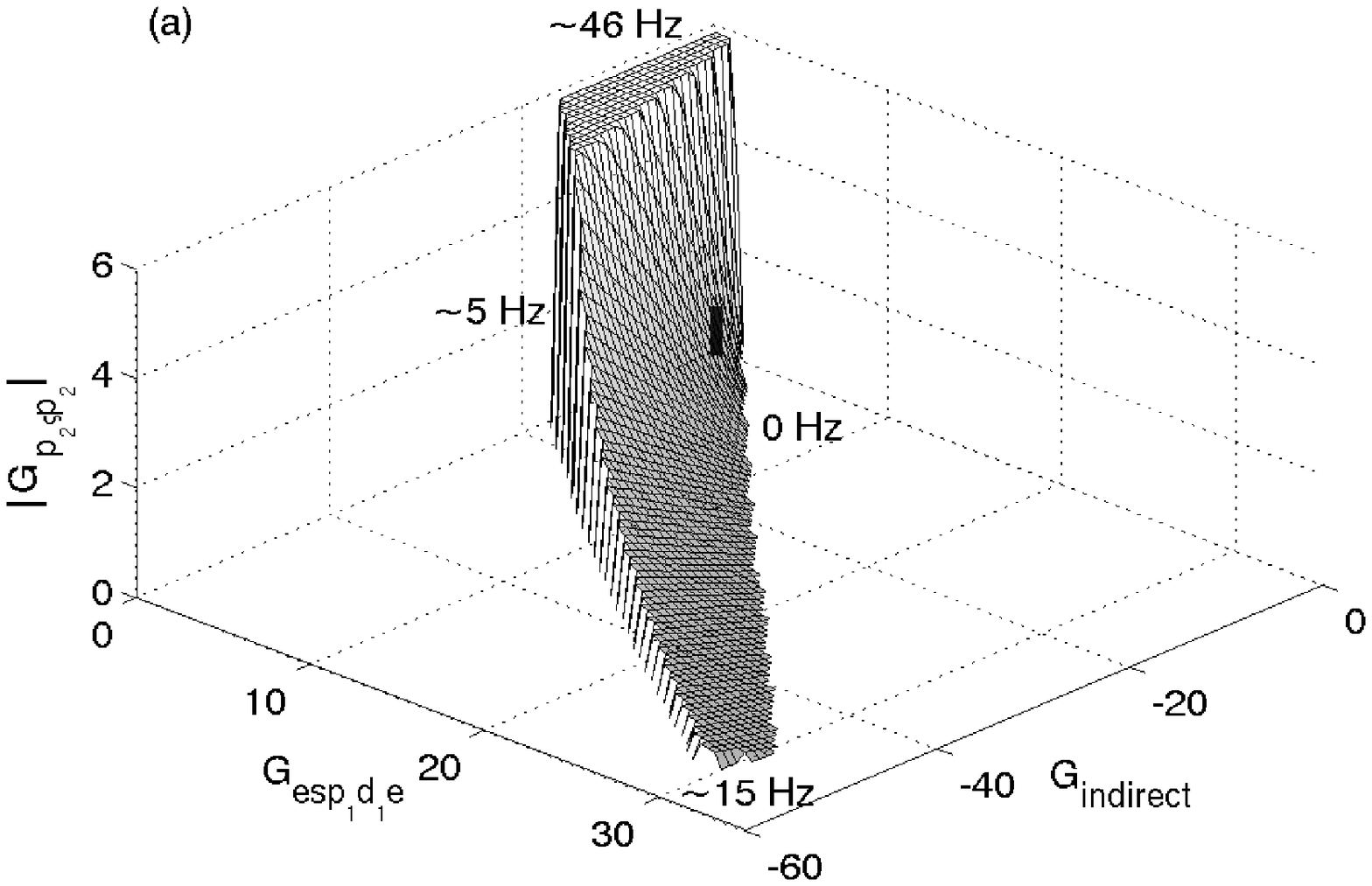}}
\subfigure{\label{fig:tent2}\includegraphics[width=220pt, height=160pt]{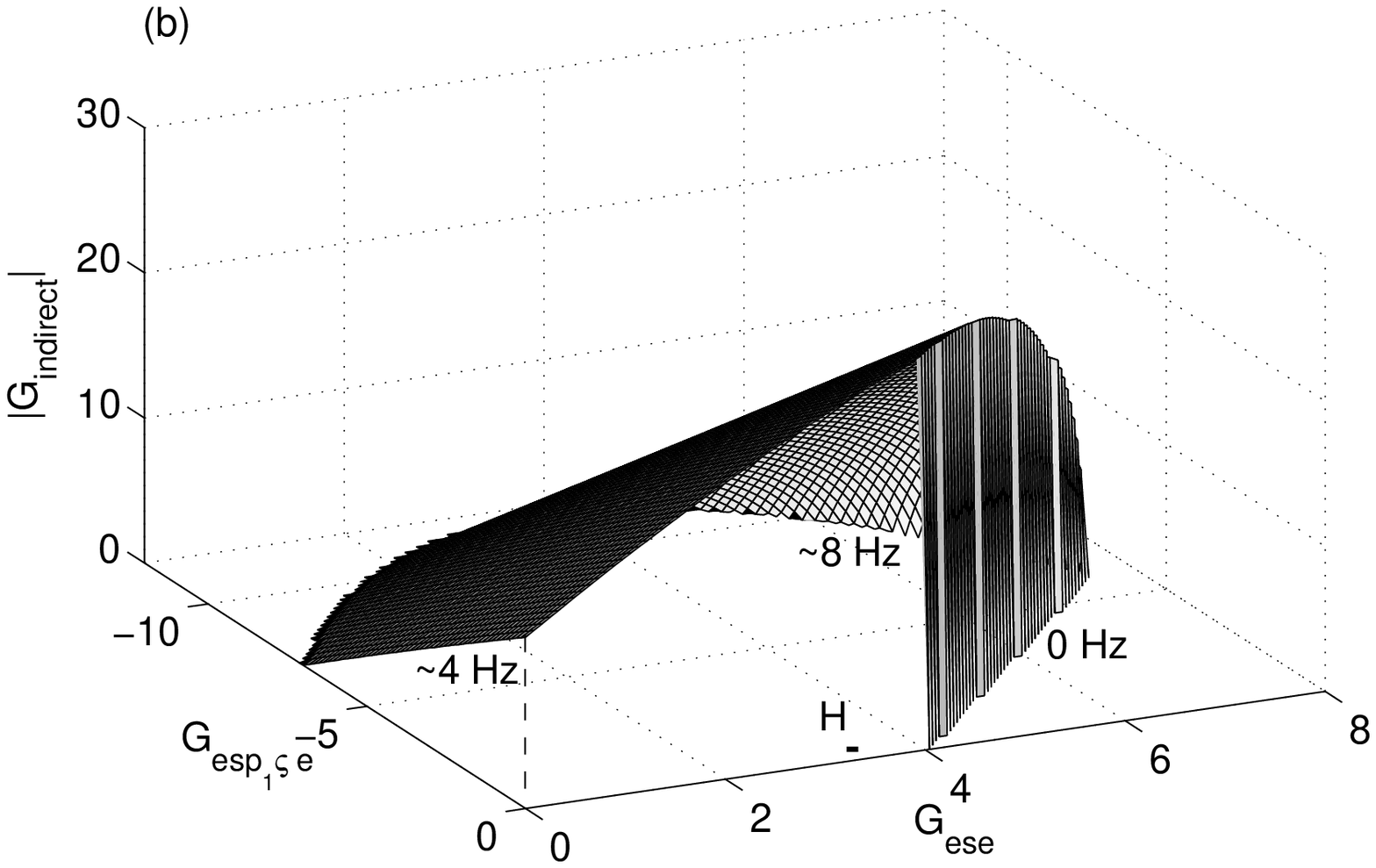}}
\caption{Regions of linear stability for the BGTCS. (a) Region parameterized by the gains of the direct and indirect pathways (where $G_{\mathrm{indirect}}=G_{esp_1\varsigma p_2d_2e}+G_{esp_1p_2d_2e}$), and the absolute value of the gain for the STN-GPe loop. The unstable frequencies are indicated for the following situations: dominance of the direct loop (0 Hz), dominance of the indirect loop ($\sim$5 Hz), a strong STN-GPe loop ($\sim$46 Hz), and both strong direct and indirect loops ($\sim$15 Hz). The top of the bar corresponds to the normal state with parameters as in Paper I. (b) Region of linear stability parameterized by the gains of the hyperdirect, corticothalamic, and indirect loops (where $|G_{\mathrm{indirect}}|=|G_{esp_1\varsigma p_2d_2e}+G_{esp_1p_2d_2e}|$). The location of the healthy state, within the stable region, is indicated by `H'. Theta instabilities arise for large gains of the indirect or hyperdirect loops, at frequencies that increase with $G_{ese}$ and $|G_{esp_1\varsigma e}|$. A 0 Hz instability occurs for large $G_{ese}$ and small $|G_{esp_1\varsigma e}|$. The dashed line indicates that the boundaries in front lie in the $G_{ese}=0$ and $G_{esp_1\varsigma e}=0$ planes. The vertical lines along the right-hand boundary are numerical artifacts.}
\end{figure}

For a given set of parameters, we can visualize the effects of changes in gains using a surface plot of the region in which the system is linearly stable. Figure \ref{fig:tent} indicates the boundary of the linearly stable region parameterized by the gains of the direct loop, the combined classic and alternative indirect loops (the latter involving the GPe-GPi/SNr projection), and the STN-GPe loop, where the other gains are held constant at the normal values obtained from the parameters in Paper I. The top of the bar corresponds to the healthy state. The left-hand boundary corresponds to an approximately 5 Hz instability that arises when the indirect pathway dominates. When $|G_{p_2\varsigma p_2}|$ becomes large, the STN-GPe loop sustains oscillations around 46 Hz. The right-hand boundary indicates that a dominant direct pathway causes the system to become unstable at 0 Hz. Finally, the small region at the front corresponds to an instability at $\sim$15 Hz when both the direct and indirect pathways are strong.

A similar diagram illustrates the dependence of stability and the frequency of theta oscillations on the gains of the hyperdirect, indirect, and corticothalamic loops [Fig. \ref{fig:tent2}]. As discussed above, theta instabilities can arise for either strong hyperdirect or indirect pathways. The analogous roles of the hyperdirect and indirect loops are further indicated by the fact that the system remains stable for larger $|G_{\mathrm{indirect}}|=|G_{esp_1\varsigma p_2d_2e}+G_{esp_1p_2d_2e}|$ (Loops 2 and 3) if $|G_{esp_1\varsigma e}|$ is relatively small. However, phase relationships, likely changes in corticostriatal connection strengths, and firing rates support the scenario in which $|G_{\mathrm{indirect}}|$ is large compared to $|G_{esp_1\varsigma e}|$. Figure 9(b) shows that the system displays an instability at 0 Hz for strong corticothalamic interactions but a weak hyperdirect pathway. The frequency of theta oscillations increases with both $G_{ese}$ and $|G_{esp_1\varsigma e}|$, ranging from about 4.2 to 7.8 Hz, which closely matches the range of tremor frequencies in parkinsonian patients. The lower frequencies of theta oscillations for smaller $G_{ese}$ (a putative index of arousal) could be related to the slowing of tremor seen with age \citep{Deuschl1996}.

\subsection{Changes in spectra with dopamine loss}\label{sec:full_spectra}

\begin{figure}[!ht]
\centering
  \includegraphics[width=420pt]{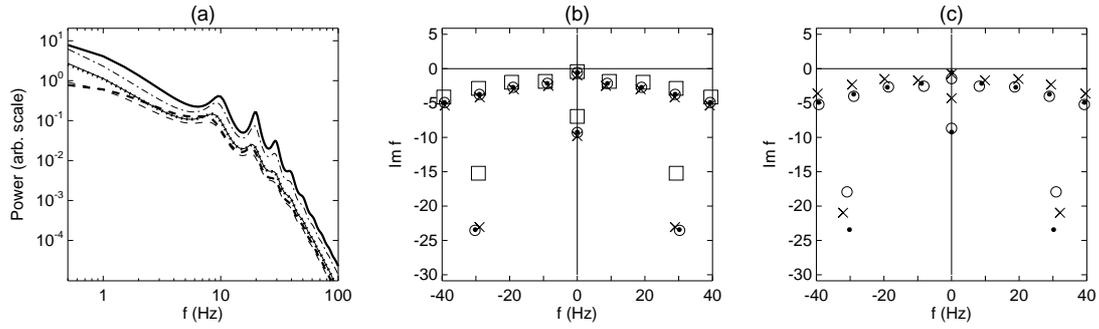}
 \caption{Linear cortical spectra and dispersion roots for mean $\phi_n$ of 10 s$^{-1}$. (a) Linear frequency spectra of the cortical signal: thick solid line, normal state with parameters as in Paper I; dotted, `reduced-SNR' state with $\theta_{d_1}=\theta_{d_2}=13$ mV, $\nu_{d_1e} = 0.4$ mV s, and $\nu_{d_2e} = 0.1$ mV s; thin dashed line, weaker direct and stronger indirect pathway, $\nu_{d_1e} = 0.5$ mV s and $\nu_{d_2e} = 1.4$ mV s; dash-dotted, reduced intrapallidal inhibition, $\nu_{p_2p_2}=-0.03$ mV s; thick solid line, $\nu_{ee}=\nu_{ie}=1.4$ mV s and $\nu_{ei}=\nu_{ii}=-1.6$ mV s to model mesocortical dopamine loss; thick dashed line, full parkinsonian state (cf. Sec. \ref{sec:changes}). Note that the spectrum for the reduced-SNR state almost coincides with that in the normal state. (b) Solutions $f=\omega/(2\pi)$ of the dispersion relation $q^2=0$ [cf. Eq. (\ref{eq:dispersion_relation})]. Proximity to the real axis determines the amplification of the signal at the corresponding frequency. Filled dots, normal state; open circles, reduced-SNR state; crosses, $\nu_{d_1e} = 0.5$ mV s and $\nu_{d_2e} = 1.4$ mV s; squares, $\nu_{p_2p_2}=-0.03$ mV s. (c) Dispersion roots for the following cases: filled dots, normal state; crosses, $\nu_{ee}=\nu_{ie}=1.4$ mV s and $\nu_{ei}=\nu_{ii}=-1.6$ mV s; open circles, full parkinsonian state.}
\label{fig:linear_spectra}
\end{figure}

We now investigate the influence of dopamine loss on frequency spectra. Using the transfer function (\ref{transfer}) we calculate the linear cortical spectrum up to a proportionality factor via $P(\omega) \propto |\phi_e^{(1)}(\omega)|^2$. The results are shown in Fig. \ref{fig:linear_spectra}(a). If parkinsonism is modeled via $\nu_{d_1e}=0.4$ mV s, $\nu_{d_2e} = 0.1$ mV s, and $\theta_{d_1}=\theta_{d_2}=13$ mV, parameters that represent a decreased striatal SNR, the cortical spectrum is virtually indistinguishable from that in the normal state. Modeling dopamine loss with $\nu_{d_1e} = 0.5$ mV s and $\nu_{d_2e} = 1.4$ mV s shifts the frequency of the alpha root from 8.9 Hz to 8.5 Hz, in line with slowed alpha peaks seen in PD patients \citep{Sinanovic2005,Soikkeli1991}. A stronger indirect and weaker direct pathway also reduces overall cortical power and relative power at 7--12 Hz from 20\% to 13\%, but increases relative power at 3--7 Hz from 9\% to 17\%. Changes in relative alpha and theta power accord with empirical findings \citep{Bosboom2006, Neufeld1994, Stoffers2007}, but absolute power is reduced in contrast to what has been found experimentally \citep{Moazami-Goudarzi2008, Tanaka2000}. Reduced intrapallidal inhibition and particularly loss of mesocortical dopamine may account for some of the experimentally observed increase in power (cf. Fig. \ref{fig:linear_spectra}). It is interesting to note that balancing the relative strengths of excitation and inhibition in the cortex also produces an increase in relative theta and beta power. Thus, our model confirms that lateral disinhibition in the cortex can account for the experimentally observed co-production of low- and high-frequency activity, which has been termed the `edge effect' \citep{Llinas2005}. The full parkinsonian state combines increased relative 3--7 Hz (23\% vs. 9\%) and decreased relative 7--12 Hz power (15\% vs. 20\%) with overall power not much lower than in the normal state (total power $\geq$ 1 Hz is 89\% of the normal value). Assuming that cortical power is slightly reduced when the striatum is significantly involved, in line with the suppressed excitability of regions normally co-activated with the striatum, or during tasks that specifically require striatal activation \citep{Monchi2004,Monchi2007}, these are realistic spectral changes. In cortical areas not directly linked with the striatum, power may be increased due to diffuse loss of dopaminergic innervation, or due to compensatory mechanisms. Finally, changes in parameters other than corticostriatal coupling strengths further slow the alpha root to 8.2 Hz in the full parkinsonian state.

The dispersion relation for the spatially uniform case is $q^2=0$ [cf. Eq. (\ref{eq:dispersion_relation})]. Its solutions up to 40 Hz are plotted in Figs. \ref{fig:linear_spectra}(b) and \ref{fig:linear_spectra}(c) for the normal and dopamine-depleted states. Each of the states represents a stable system, since all roots are found in the lower half plane. Increased relative theta power corresponds to a smaller distance between the least stable roots on the imaginary axis, which `pulls' the alpha roots to lower power and frequency. These root locus diagrams also show that reduced intracortical inhibition and a lower GPe threshold potential enhance gamma-band power around 35 Hz in the STN-GPe network. The frequency of this rhythm goes up as the corresponding roots move closer to the real axis, explaining the higher frequency ($\sim$46 Hz) of limit cycles in the STN-GPe loop.

\begin{figure}[!ht]
\centering
\includegraphics[width=420pt]{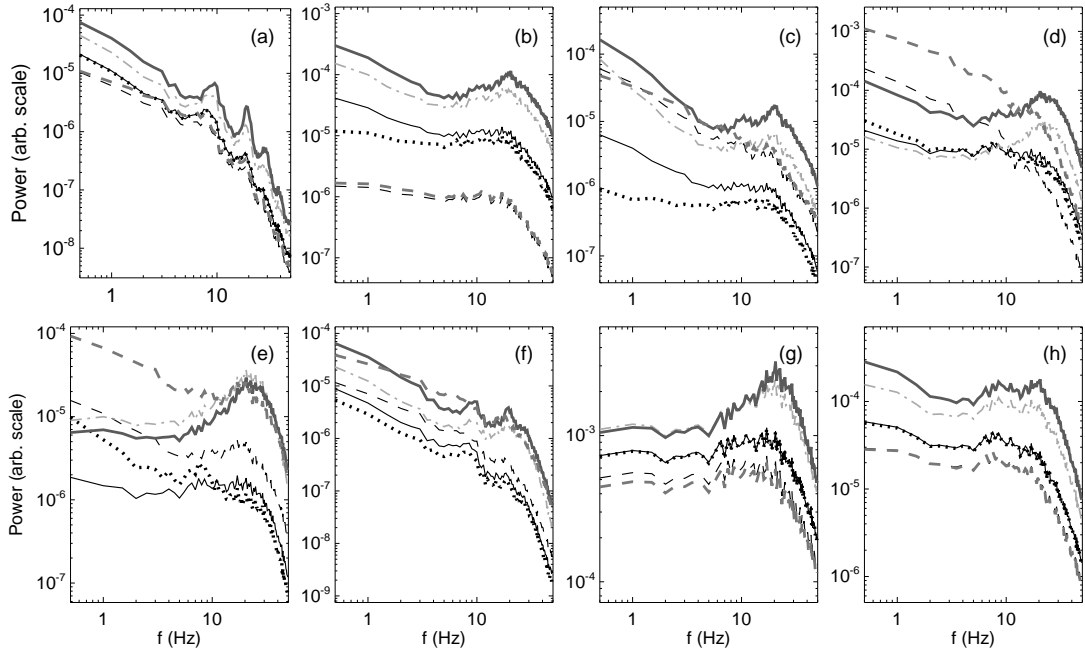}
\caption{Spectra obtained by numeric integration and averaging the Fourier transforms of 60 successive 2-s epochs. Input consisted of Gaussian white noise with mean 10 s$^{-1}$ and standard deviation 2 s$^{-1}$. Thin solid lines, normal state; dotted, `reduced-SNR' state with $\theta_{d_1}=\theta_{d_2}=13$ mV, $\nu_{d_1e} = 0.4$ mV s, and $\nu_{d_2e} = 0.1$ mV s; thin dashed lines, state with $\nu_{d_1e} = 0.5$ mV s, $\nu_{d_2e} = 1.4$ mV s; dash-dotted, with reduced intrapallidal inhibition, $\nu_{p_2p_2} = -0.03$ mV s; thick solid lines, with $\nu_{ee}=\nu_{ie}=1.4$ mV s and $\nu_{ei}=\nu_{ii}=-1.6$ mV s; thick dashed lines, full parkinsonian state (cf. Sec. \ref{sec:changes}). (a) $\phi_e$; (b) $\phi_{d_1}$; (c) $\phi_{d_2}$; (d) $\phi_{p_1}$; (e) $\phi_{p_2}$; (f) $\phi_{\varsigma}$; (g) $\phi_s$; (h) $\phi_r$.}
\label{fig:nonlinear_spectra}
\end{figure}

Spectra obtained by numeric integration of the full nonlinear equations are shown in Fig. \ref{fig:nonlinear_spectra}. These were computed by averaging the Fourier transforms of 60 consecutive 2-s epochs for a Gaussian white noise input with mean 10 s$^{-1}$ and standard deviation 2 s$^{-1}$. The main results of parameter changes mimicking a reduced striatal SNR are lower-amplitude fluctuations in D1 and D2 cells, and increased low-frequency power in the GPe. The parameters $\nu_{d_1e}=0.5$ mV s and $\nu_{d_2e}=1.4$ mV s lead to decreased corticothalamic power, and increased power in all basal ganglia populations except D1. This accords with the enhanced fluctuations in the responses of the basal ganglia, and reduced fluctuations in thalamocortical responses to ongoing inputs seen in Sec. \ref{sec:ongoing_responses}. Furthermore, these parameter changes amplify relative power at 15--25 Hz in the cortex, D1 neurons, and the STN, in line with increased beta coherence between cortex and STN in PD patients \citep{Brown2001,Marsden2001}. Lateral disinhibition of the GPe ($\nu_{p_2p_2}=-0.03$ mV s) enhances fluctuations in all populations, and particularly activity around 20--30 Hz in both pallidal segments and the STN. Weaker cortical interactions ($\nu_{ee}=\nu_{ie}=1.4$ mV s, $\nu_{ei}=\nu_{ii}=-1.6$ mV s) cause similar changes, which are more pronounced in all components except the GPe than changes caused by reduced intrapallidal inhibition. Figure \ref{fig:nonlinear_spectra} shows that the full parkinsonian state is accompanied by increased relative 3--7 Hz power throughout the BGTCS compared to the healthy state. Relative power at 15--25 Hz is enhanced in the cortex (5.2\% vs. 4.0\%), D1 neurons (27\% vs. 25\%), and the STN (12\% vs. 5\%), and decreased in the remaining populations, partly due to the increase in theta power. As a fraction of power $\geq 7$ Hz, 15--25 Hz activity is enhanced also in the GPe (38\% vs. 36\%).

\subsection{The paradox of GPe lesion}\label{sec:GPe_lesion}

\begin{figure}[!ht]
\centering
  \includegraphics[width=420pt]{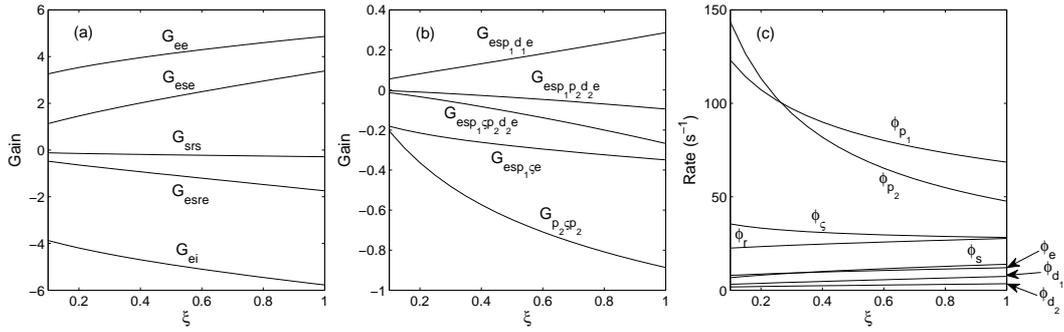}
  \caption{Effects of multiplying $\nu_{p_1p_2}, \nu_{\varsigma p_2},$ and $\nu_{p_2p_2}$ by a factor $\xi$ to mimic GPe lesion. (a) Corticothalamic gains. (b) Gains involving the basal ganglia. (c) Steady-state firing rates vs. $\xi$.}
\label{fig:GPe_lesion}
\end{figure}

\citet{Leblois2006} did not include the indirect pathway in their model partly because of the finding by \citet{Soares2004} that GPe lesion in the monkey does not lead to parkinsonian motor symptoms or altered activity patterns in the GPi. The authors concluded that this invalidates the indirect loop as a candidate for the origin of synchronous oscillations and motor symptoms in PD. We modeled GPe lesion by reducing the absolute values of all gains emanating from the GPe, $\nu_{p_1p_2}, \nu_{p_2p_2},$ and $\nu_{\varsigma p_2}$, multiplying each by a factor $\xi$ between 0.1 and 1 with results shown in Fig. \ref{fig:GPe_lesion}. Although corticothalamic rates are reduced, while pallidal and STN rates are increased by GPe lesion in our model, a reduction in all the gains implies that oscillations are damped rather than enhanced in the indirect, hyperdirect, and STN-GPe loops. However, as we saw in the previous sections, strengthening of the indirect pathway by either increasing $\nu_{d_1e}$ or $\nu_{d_2e}$ can lead to slow oscillations in this loop. Thus, a putative lack of parkinsonian signs following GPe lesion does not preclude the involvement of the indirect pathway in the generation of synchronous oscillations and motor symptoms. In fact, \citet{Chesselet1996} noted that GPe lesion does not necessarily reflect what happens with nigrostriatal degeneration, precisely because it does not reproduce changes elsewhere, such as in STN or GPi. Besides, the negative result of \citet{Soares2004} may be related to a relatively small extent or to the location of the lesions, since \citet{Zhang2006} did report worsened akinetic symptoms after GPe lesions in MPTP-treated rhesus monkeys. This matches the smaller value of $|G_{ee}+G_{ei}|$ with reduced GPe output in our model (0.6 for $\xi=0.1$ vs. 0.9 in the healthy state, if impaired cortical inhibition is linked with akinesia (cf. Sec. \ref{sec:gains}).

\section{Summary and discussion}

Using a physiologically based mean-field model of the basal ganglia-thalamocortical system (BGTCS), we have reproduced many of the electrophysiological correlates of Parkinson's disease (PD) with parameter values in the healthy and parkinsonian states estimated from known physiology. The analysis builds on a companion paper \citep{vanAlbada2008}, which examined the effects of nigrostriatal dopamine depletion on firing rates. In that paper we showed that an increase in the cortical connection strength to striatal neurons expressing the D2 class of dopamine receptor, possibly along with a reduced connection strength to D1-expressing striatal neurons, leads to increased rates of the striatum, STN, and output nuclei, and decreased GPe and thalamic rates, in good agreement with many experiments. On the other hand, simultaneous decreases in corticostriatal connection strengths and striatal firing thresholds, chosen to mimic a reduced striatal signal-to-noise ratio,
had little effect on rates. We note that these results do not bear directly on the role of dopamine as a contrast enhancer, since these particular parameter changes may not adequately capture a reduced signal-to-noise ratio in the striatum as a whole, and dopamine may enhance differential responses to strong and weak inputs in some striatal neurons but not others. Disinhibition of the cortex due to impaired local dopaminergic innervation helped normalize the cortical firing rate, as seen experimentally. Other changes secondary to nigrostriatal damage, including lower GPe and STN firing thresholds and weaker intrapallidal inhibition, could account for the comparatively large increase in STN rate and conflicting findings on changes in GPe rate. In the present work we have investigated the dynamical implications of the above ways of modeling dopamine depletion. The wide range of phenomena accounted for suggests that the model provides a physiologically realistic representation of the mean-field dynamics of the BGTCS. Our main findings are listed below.\\

\noindent \emph{Outcomes of model}\\

(i) Reduced availability of cortical dopamine may contribute to bradykinesia in PD patients by attenuating intracortical excitation by pyramidal cells, and especially inhibition by interneurons \citep{Gulledge2001,Gao2003,Thurley2008}, resulting in a smaller difference between excitatory and inhibitory gains. This causes more gradual changes in cortical activity, in agreement with slowed motor unit responses to transcranial magnetic stimulation in PD patients \citep{Kleine2001}, and slowed responses of motor cortex neurons in monkeys treated with the neurotoxin 1-methyl-4-phenyl-1,2,3,6-tetrahydropyridine (MPTP) \citep{Doudet1990}. Also, the silent period after suprathreshold stimulation of the cortex during muscle contraction is shortened in parkinsonian patients \citep{Cantello1991}, which is reversed by dopaminergic medication \citep{Priori1994}. Since the silent period is considered to be caused by the activation of GABAergic interneurons \citep{Fuhr1991}, this indicates the importance of impaired cortical inhibition in bradykinesia. Our model suggests that some of these changes are also explained by nigrostriatal dopamine loss, which can decrease cortical gains, and the difference between the strengths of intracortical excitation and inhibition, due to a shift in balance from the direct to the indirect pathway. 

Other explanations for bradykinesia/akinesia have been put forward, such as overactivity of GPi and/or STN projections to the PPN \citep{Munro-Davies1999}, which in turn innervates the pontine and medullary reticulospinal systems \citep{Jackson1983}. In addition, rigidity, muscle weakness, tremor, slowing of thought, and compensation for loss of movement accuracy may all contribute to bradykinesia \citep{Berardelli2001}. Thus, other factors are likely to complement the mechanism proposed here.

(ii) An increase in the cortico-D2 connection strength with or without a weaker cortico-D1 projection leads to realistic changes in responses to transient stimuli. More vigorous model responses are observed in the STN and the output nuclei, in accord with experiments \citep{Bergman1994,Filion1988,Wichmann1994}. As also observed experimentally \citep{Magill2001,Zold2007}, the inhibitory phase of a triphasic early GPe model response becomes more pronounced as the indirect loop gains in strength. Changes in D1 and D2 responsiveness are in line with recorded effects of dopamine on the sensitivity of striatal medium spiny neurons to cortical inputs: D1 responses are attenuated and D2 responses are amplified with dopamine loss, as seen in experiments \citep{Cepeda1998,Hernandez-Lopez1997,Hsu1995,Kiyatkin1996,Levine1996,Nicola2000,Toan1985,Umemiya1997}.

(iii) The indirect basal ganglia pathway can sustain $\sim$5 Hz oscillations if dopamine depletion leads to greater cortical influence on D2-expressing striatal neurons with or without reduced influence on D1-expressing neurons. The fact that modulation of the direct pathway need not be opposite to that of the indirect pathway underlines that a complete separation of D1 and D2 neurons is not required to obtain this result. Limit cycle oscillations around 5 Hz only appear when the indirect loop is very strong compared to thalamocortical interactions, which is unlikely to happen in the system as a whole and confines oscillations to subcircuits. This scenario is supported by the limited proportion of tremor cells recorded in components of the BGTCS in parkinsonian humans and animals \citep{Bergman1994,Lemstra1999,Levy2001, Wichmann2006}, and the high firing rate of oscillatory cells compared to non-oscillatory cells \citep{Dejean2008,Levy2000,Levy2001}. As explained in the Introduction, central oscillations need not directly determine the frequency of parkinsonian tremor, since cortical and peripheral factors appear to play a modulatory role. However, the frequency of oscillations in the indirect loop predicted by our model is close to the $\sim$4--6 Hz of parkinsonian rest tremor \citep{Deuschl1989}, as well as to measured rhythms in the basal ganglia, cortex, and thalamus \citep{Bergman1994, Magnin2000, Raz2000, Volkmann1996, Wang2005, Wichmann1999}. A relatively high level of arousal is required for limit cycle oscillations to appear, possibly helping to explain why tremor is exacerbated by anxiety and stress \citep{Deuschl2001, Zesiewicz2001}. 

(iv) The phase relationships between the basal ganglia components oscillating at $\sim$5 Hz accord with those measured by \citet{Walters2007} for $\sim$1 Hz oscillations in anesthetized rats with dopaminergic lesions. Furthermore, \citet{Zold2007b} reported a population of GP (the rodent homolog of GPe) neurons whose firing rate was modulated in antiphase to cortical slow waves and striatal activation in parkinsonian rats under anesthesia, whereas GP activity was in phase with cortex and striatum in healthy anesthetized rats. This suggests that parkinsonian slow oscillations are relayed via the striatopallidal axis rather than via the hyperdirect pathway through the STN, since the GPe is inhibited by the striatum, but excited by the STN. If these results generalize to PD patients, this is strong evidence for an origin of parkinsonian theta oscillations in the indirect loop. \citet{Leblois2006} proposed that the hyperdirect loop is the main substrate of parkinsonian oscillations based on a neuronal network model of the direct and hyperdirect loops. This would imply that GPe activity would be modulated in phase with STN activity, being driven by excitatory input from this nucleus, contrary to experimental observations. Moreover, an origin in the hyperdirect loop was suggested based on the assumption that firing rates are virtually unchanged, which does not appear to be generally applicable (see Paper I). In our model, a source of parkinsonian oscillations in the hyperdirect loop is incompatible with increased striatal rates and a decreased GPe rate, often measured in animal models of PD \citep{Boraud1998, Chen2001, Filion1991, Heimer2002, Kish1999, Pan1988, Walters2007}. In contrast, our model predicts both realistic changes in firing rates and oscillations in the indirect loop simultaneously for a stronger cortico-D2 connection. Moreover, striatal involvement in parkinsonian rhythms is strongly implied by the dense dopaminergic innervation of this nucleus, which is expected to powerfully modulate corticostriatal connection strengths.

(v) In the model, oscillations around 20 Hz that are normally largely confined to corticothalamic circuits are enhanced and spread to the basal ganglia (particularly to the STN) when the indirect loop becomes strong due to nigrostriatal degeneration, while cortex and thalamic relay nuclei are also relatively tightly coupled. Such enhanced $\sim$20 Hz oscillations have been found in many experimental studies of parkinsonism \citep{Brown2001, Brown2005, Fogelson2006, Gatev2006, Levy2002, Sharott2005}. Cortical activity has been shown to be phase-advanced with respect to STN activity at $<$30 Hz, corroborating a corticothalamic origin of beta oscillations \citep{Marsden2001,Williams2002}. Depending on the relative gains of the corticothalamic and indirect loops, model rhythms at $\sim$20 Hz may be stronger or weaker than $\sim$5 Hz rhythms, and limit cycles may therefore show a combination of these frequencies, also in accord with experiments \citep{Levy2001}.

Enhanced beta oscillations in the basal ganglia may either be a side effect of other pathological changes, or be directly related to parkinsonian symptoms, as suggested by a number of studies. For instance, \citet{Silberstein2003} found a larger percentage of local field potential activity at 11--30 Hz in the pallidum of untreated PD patients compared to treated ones. Moreover, a study in which 20 Hz stimulation of the STN slowed finger-tapping rates supported a relationship between beta oscillations and akinesia/bradykinesia \citep{Chen2007}. A connection between beta oscillations in the basal ganglia and akinetic symptoms is further supported by the fact that the oscillations are attenuated before and during voluntary movement \citep{Kuhn2004}, as is 15--30 Hz coherence between motor cortex and muscles \citep{Farmer2002}. On the other hand, the low-pass filter characteristics of the pallidal-cortex-muscle axis suggest that beta oscillations may not strongly influence movement \citep{Rivlin-Etzion2008}, and 20 Hz synchronization in the STN was found to correlate with motor improvement after dopamine replacement therapy, rather than with the initial severity of symptoms \citep{Weinberger2006}. Thus, the precise relation between beta oscillations and parkinsonian symptoms remains to be elucidated.

(vi) Our model predicts a resonant frequency in the gamma band ($>$30 Hz) for the loop formed by STN and GPe, in accord with modeling results of \citet{Humphries2006}. Alterations in this network may therefore be responsible for enhanced gamma power in the STN observed in PD patients on levodopa \citep{Fogelson2005}, particularly in relation to movement \citep{Cassidy2002}. More recently, such gamma activity was also recorded in patients after overnight withdrawal from antiparkinsonian medication \citep{Trottenberg2006}. In PD patients, STN activity has been found to lead cortical activity in the gamma band \citep{Williams2002}, supporting our finding of an origin in the STN region.

(vii) Stronger cortico-D2 coupling increases the amplitude of fluctuations in firing rates across the spectrum in D2-expressing striatal neurons, GPe, STN, and the output nuclei, suggesting an increase in synchronization between individual cells. This applies especially to the GPe, where the firing rate is decreased, since increased rates may enhance power without changing the degree of synchronization. However, power at low frequencies is amplified by a factor of up to 40 in the output nuclei, whereas the increase in average rate from 69 s$^{-1}$ to 112 s$^{-1}$ only accounts for an approximate factor of 2.7 if the standard deviation is taken to be proportional to the average rate. STN power is also more amplified than accounted for by its higher average rate in the parkinsonian state. This strongly implies that synchronization is enhanced, in concordance with experimental studies \citep{Goldberg2004,Hammond2007,Heimer2002, Levy2000,Morris2005}. \citet{Levy2002} proposed that only oscillatory cells are synchronized in parkinsonism, although this study may have failed to detect fluctuations at the population level, and did not compare with healthy subjects. Our results suggest that dopamine loss increases synchronization across the whole spectrum, rather than only in peaks, thus including cells whose activity is not modulated at one of the main resonant frequencies of the system. This result requires further experimental verification.

(viii) Dominance of the indirect pathway increases relative theta power, decreases relative alpha power, and shifts the alpha peak of the electroencephalogram (EEG) to lower frequencies, in agreement with empirical findings \citep{Bosboom2006, Neufeld1994, Sinanovic2005,Soikkeli1991,Stoffers2007}. Changes secondary to nigrostriatal dopamine loss, such as reduced intrapallidal inhibition, a stronger striato-GPe projection due to enhanced release of GABA, and lower GPe and STN threshold potentials, may also contribute to these spectral changes by further increasing the gain of the indirect loop. More pronounced EEG frequency reduction in demented patients may be partly explained by additional changes in cholinergic and noradrenergic signaling \citep{Buzsaki1988,Candy1983,Cash1987,Dubois1983, Metherate1995}. Total EEG power is decreased by a stronger indirect pathway in the model, in accord with its inhibitory effect on the thalamus and cortex. However, EEG power has been reported to be increased in PD \citep{Moazami-Goudarzi2008, Tanaka2000}. Loss of intrinsic cortical dopamine, reduced intrapallidal inhibition, and a lower GPe firing threshold may partly normalize EEG power that is reduced by a dominant indirect pathway. Considering that cortical responses are suppressed in PD in tasks or regions with significant involvement of the striatum \citep{Monchi2004,Monchi2007}, we do not expect these changes to increase EEG power beyond the normal level. Rather, diffuse loss of mesocortical dopamine leading to cortical disinhibition \citep{Mattay2002}, and compensatory changes in areas not directly connected to the basal ganglia \citep{Samuel1997}, may cause the observed EEG amplification, whereas we expect EEG fluctuations in areas strongly connected with the basal ganglia to be diminished.\\

\noindent \emph{Wider context and future directions}\\

The above results were obtained using a small number of variations around a single set of parameters representing the healthy state. Thus, the same axonal, dendritic, and synaptic delays that give realistic responses to transient cortical stimuli, a resonant frequency of the STN-GPe loop in the gamma band, and an alpha peak frequency around 9--10 Hz, predict frequencies around 5 Hz and 20 Hz for oscillations in the parkinsonian state. Moreover, the same parameter changes that yield plausible firing rate changes with nigrostriatal damage lead to increased relative theta power, decreased relative alpha power, and a lower alpha peak frequency, all in accord with experiments. This provides firm support for the proposed mechanisms, and strongly suggests that combinations of parameter values required to obtain these results were chosen in physiologically realistic ranges.

The debate on the substrate of parkinsonian oscillations has been fueled by paradoxical results concerning changes in GPi and STN activity and motor behavior following GPe lesion. \citet{Soares2004} observed a lack of parkinsonian symptoms or oscillatory bursting after GPe lesion, prompting \citet{Leblois2006} to reject the indirect pathway as a possible source of parkinsonian oscillations. However, despite the generally decreased firing rate of the GPe in parkinsonism, nigrostriatal and GPe lesions can have notably different effects on firing patterns in the BGTCS. This is apparent in our model from the fact that weaker efferent projections from the GPe lower all gains, rendering the system more stable and damping oscillations. Therefore, the supposed lack of parkinsonian symptoms following GPe lesion does not preclude the possibility that tremor rhythms arise in the indirect basal ganglia circuit. In fact, \citet{Soares2004} remarked that the involvement of the GPe is supported by the observation that oscillatory cells in the STN and GPi were slightly less numerous in GPe-lesioned animals than in intact animals. Furthermore, our model predicts that damage to the GPe impairs intracortical inhibition more than excitation, analogous to nigrostriatal damage, which matches the finding by \citet{Zhang2006} that GPe ablation does exacerbate the parkinsonian symptoms of akinesia/bradykinesia in rhesus monkeys previously treated with MPTP.

The different effects of GPe lesion and nigrostriatal lesions that reduce GPe activity exemplify the more general rule that a single structure can contribute to a variety of phenomena, depending on its connections and the parameter values of the system. Single circuits can also support different types of activity, as evidenced for instance by the different frequencies of oscillations in the indirect loop depending on the relative values of corticothalamic and indirect loop gains. This example immediately makes it clear that multiple structures can also contribute to a single effect, underlining the importance of including circuit interactions in models of neural systems rather than focusing only on localized `pacemakers', as has sometimes been done in the past. An aspect of such models that is perhaps less often considered is that certain activity patterns can be sustained by multiple interacting circuits. An example was given in Sec. \ref{sec:full_stability}, where oscillations were derived arising from an interplay between corticothalamic circuits and the indirect loop, rather than in any circuit separately.

It is significant to note that bursting activity is not required to account for increased theta and beta oscillations and EEG slowing. Accordingly, bursting has gradually been deemphasized in the literature as an explanation for parkinsonian oscillations and symptoms, since it appears to be particularly prevalent during dyskinesias \citep{Lee2001, Silberstein2003}, treatment with the dopamine agonist apomorphine increases aperiodic bursting in STN and GPi \citep{Levy2001}, and bursts around tremor frequency give way to slower rhythmic bursts during voluntary movement \citep{Rodriguez-Oroz2001}. Nevertheless, calcium spike bursts in the relay nuclei may contribute to enhanced thalamocortical theta coherence \citep{Sarnthein2007}, and to overproduction of beta and gamma activity in the cortex via the edge effect \citep{Llinas2005,Moazami-Goudarzi2008}. Therefore, taking into account thalamic bursting in our model may help explain increases in EEG power in the theta, beta, and gamma bands. It would also be relevant to include bursting properties of STN \citep{Beurrier1999, Bevan2000} and GPe \citep{Nambu1994} cells, which may modify the amplitude, frequency, and timing of rhythms predicted by our model. We aim to modify our model by including distinctive electrophysiological properties of STN, GPe, and thalamic bursting neurons in future work.

In addition to the functional interactions included in our model so far, there are projections from the GPe to the TRN \citep{Gandia1993,Hazrati1991} and to striatal interneurons \citep{Bevan1998}, from the thalamic parafascicular nucleus to the STN \citep{Hassani1997, Mouroux1995}, and between the PPN and all elements of the BGTCS \citep{Hammond1983, Jackson1983,Lavoie1994a,Orieux2000}. Studies in rats with nigrostriatal lesions have suggested that excitatory projections from the PPN are partly responsible for STN hyperactivity \citep{Breit2006}, despite the loss of neurons from this region in PD \citep{Zweig1989}, and the induction of akinesia by PPN lesion in otherwise healthy primates \citep{Munro-Davies1999,Pahapill2000}. Some of these projections may be included in future work. Instead of more detailed modeling, an alternative approach would be to simplify the current model to extract those features essential for explaining phenomena such as changes in firing rates, theta, beta, or gamma rhythms, or trends in EEG spectra. Simplified models could be analyzed more systematically and provide more robust parameter estimates, although these estimates would be less easily related to the underlying complex physiology. Thus, both more detailed and sparser modeling can provide new information complementing the predictions of the current model.

\section*{Acknowledgments}
The authors are grateful to J.M. Clearwater and C.J. Rennie for help with some figures. This work was supported by an Endeavour International Postgraduate Research Scholarship, an International Postgraduate Award, and the Australian Research Council. 

\singlespacing

\end{document}